\documentclass[preprint2]{aastex6}

\def\apjs{{\rm ApJS}}                  

\def\ie{{\it i.e.} }                    
\def\a4{\hsize 17.0cm \vsize 25.cm}

\shorttitle{Rapidly cooling shocked stellar winds.}
\shortauthors{ W\"unsch et al.}

\begin{document}

\title{The formation of secondary stellar generations in massive young star clusters from 
rapidly cooling shocked stellar winds}
\author{R. W\"unsch, J. Palou\v s}
\affil{Astronomical Institute, Academy of Sciences of the Czech
Republic, Bo\v{c}n\'\i\ II 1401, 141 31 Prague, Czech Republic}
\author{G. Tenorio-Tagle}
\affil{Instituto Nacional de Astrof\'\i sica Optica y
Electr\'onica, AP 51, 72000 Puebla, M\'exico}
\author{S. Ehlerov\' a}
\affil{Astronomical Institute, Academy of Sciences of the Czech
Republic, Bo\v{c}n\'\i\ II 1401, 141 31 Prague, Czech Republic}

\begin{abstract}

We study a model of \emph{rapidly cooling shocked stellar winds} in young
massive clusters and estimate the circumstances under which secondary star
formation, out of the reinserted winds from a first stellar generation (1G), is
possible. We have used two implementations of the model: a highly idealized
computationally inexpensive spherically symmetric semi-analytic model, and a
complex three-dimensional radiation-hydrodynamic simulations, and they are in a
good mutual agreement. The results confirm our previous findings that in a
cluster with 1G mass $10^7$\,M$_\odot$ and half-mass radius $2.38$\,pc, the
shocked stellar winds become thermally unstable, collapse into dense gaseous
structures that partially accumulate inside the cluster, self-shield against
ionizing stellar radiation and form the second generation (2G) of stars. We have
used the semi-analytic model to explore a subset of the parameter space covering
a wide range  of the observationally poorly constrained parameters: the heating
efficiency, $\eta_\mathrm{he}$, and the mass loading, $\eta_\mathrm{ml}$. The
results show that the fraction of the 1G stellar winds accumulating inside the
cluster can be larger than $50$\% if $\eta_\mathrm{he} \lesssim 10$\% which
is suggested by the observations. Furthermore, for low $\eta_\mathrm{he}$, the
model provides a self-consistent mechanism predicting 2G stars forming only in
the central zones of the cluster. Finally, we have calculated the accumulated
warm gas emission in the H30$\alpha$ recombination line, analyzed its velocity
profile and estimated its intensity for super star clusters in interacting
galaxies NGC4038/9 (Antennae) showing that the warm gas should be detectable
with ALMA.

\end{abstract}


\section{Introduction}
\label{sec:intro}


Young massive clusters with masses $M_\mathrm{SC} = 10^5$ -- $10^7$\,M$_\odot$
observed in nearby starburst galaxies \citep[e.g.][and references
therein]{2010ARA&A..48..431P, 1995AJ....109..960W, 2007ApJ...668..168G,
2008A&A...489.1091M, 2005ApJ...619..270M, 2007ApJ...671..358W} include high
numbers of massive stars concentrated within a small volume (radius of a few
parsecs). Winds of these stars collide with each other and convert their kinetic
energy into heat resulting in a hot gas filling most of the cluster interior.
The high thermal pressure of this gas drives a star cluster wind that becomes
supersonic as it expands into the surrounding medium. Considering that
global parameters of stellar winds vary on a time-scale which is longer than the
wind crossing time through the cluster, \citet{1985Natur.317...44C} found basic
properties of the stationary (i.e. time independent) cluster wind by solving
adiabatic, spherically symmetric hydrodynamic equations.

\citet{2003ApJ...590..791S} found that when radiative cooling of the hot shocked
wind is taken into account, the stationary star cluster wind solution does not
exist if the total cluster mechanical luminosity, $L_\mathrm{SC}$, exceeds the
so called \emph{critical luminosity}, $L_\mathrm{crit}$. The critical luminosity
is a function of other cluster parameters, e.g. it can be shown that it is
directly proportional to the star cluster radius $R_\mathrm{SC}$
\citep{2004ApJ...610..226S}. Since $L_\mathrm{SC}$ is directly proportional to
the star cluster stellar mass, $M_\mathrm{SC}$, the condition for the existence
of the stationary solution can be also formulated as an upper limit for the
cluster compactness $C \equiv M_\mathrm{SC}/R_\mathrm{SC}$ \citep[cnf.
to][]{2016A&A...587A..53K}. \citet{2005ApJ...620..217T} hypothesized that in
clusters with $L_\mathrm{SC} > L_\mathrm{crit}$, the mass reinserted by massive
stars (i.e. shocked stellar winds and supernova ejecta) accumulates in the
cluster interior and feeds secondary star formation in situ.
\citet{2007ApJ...658.1196T} confirmed the mass accumulation by 1D hydrodynamic
simulations and found that such  clusters present two qualitatively different
regions (hence the solution was named \emph{bimodal}) separated by the so called
\emph{stagnation radius}, $R_\mathrm{st}$: the mass inserted below it
accumulates inside the cluster while the mass inserted above it leaves the
cluster in a form of a cluster wind for which the stationary solution can be
found. The process of mass accumulation in star clusters with a  bimodal
solution was explored by \citet{2008ApJ...683..683W} who ran 2D hydrodynamic
simulations and found that parcels of the hot gas below the stagnation radius
cool rapidly from $\sim 10^7$\,K down to the minimum allowed temperature
$10^4$\,K (motivated by the assumption that the gas is ionized by stellar 
radiation) and are subsequently compressed by the surrounding hot gas until they
reach pressure equilibrium as dense warm clumps. The bimodal solution was
further studied by \citet{2010ApJ...708.1621T} who estimated the shapes of
recombination line profiles from 2D simulation, \citet{2010ApJ...711...25S} who
applied it to high-redshift SCUBA galaxies with extremely high star formation
rates, \citet{2010ApJ...716..324H, 2013ApJ...766...92H} who applied it to
galactic nuclear star clusters, \citet{2011ApJ...740...75W} who calculated the
time evolution of  clusters evolving in the bimodal regime for the whole period
of the existence of massive stars, and \citet{2013ApJ...778..159T} who included
cooling due to dust produced by supernovae.

This work follows up on two of our previous papers. In
\citet{2013ApJ...772..128P} we studied the properties of star cluster winds
produced by sources representing first generation (1G) stars, distributed
spatially according to a generalized Schuster function
\citep{1964ZA.....60...43L,1998SerAJ.158...15N}. This is a more realistic
stellar density profile of a cluster than the top-hat function (or even
distribution of sources) used in previous works. In \citet{2014ApJ...792..105P}
we estimated the conditions under which the accumulated warm gas self-shields
against the ionizing stellar EUV radiation (with photon energies above
$13.6$\,eV) and cools below $10^4$\,K to form a second generation (2G) of stars.
Here we model stellar clusters with a first generation of stars represented by a
smooth distribution of mass and energy sources which follow the Schuster profile.
We include radiative cooling of the gas, ionizing radiation of massive stars,
and the gravitational field from the first stellar generation. We combine 3D
radiation-hydrodynamic simulations (to calculate three models with high accuracy)
and 1D semi-analytic models (to explore a larger parameter space). The
simulations include an approximate model of star formation implemented through
sink particles, the gas self-gravity, and gravity from sink particles. We
concentrate on the first $3.5$\,Myr of the cluster evolution, i.e. before
supernovae start to explode. This is due to our smooth insertion of  mass and
energy which cannot represent well discrete events such as SN explosions. (Note
however that most groups have inserted SNe in this way, assuming a time averaged
energy and mass input as inferred from the Starburst99 synthesis code
\citep{1999ApJS..123....3L}). Furthermore, the SN ejecta may be enriched by a
non-negligible amount of dust, which is an agent capable of enhancing the
cooling of the hot gas \citep{2013ApJ...778..159T} and this is not yet
implemented in our model. The cluster evolution during the SN period will be
described in a forthcoming paper (Je\v{r}\'abkov\'a, in prep.).

Motivated by the observations, we introduced two additional parameters: the
\emph{heating efficiency} $\eta_\mathrm{he}$ and the \emph{mass loading}
$\eta_\mathrm{ml}$. The first one indicates the fraction of the mechanical
energy of stellar winds that is transformed into thermal energy of the hot
shocked gas inside the cluster. This may help to solve the discrepancy between
the observed and the predicted X-ray luminosity of HII regions associated with
young massive clusters \citep[see e.g.][and the references
therein]{2014MNRAS.442.2701R}. \citet{2010ApJ...711...25S} determined the
heating efficiency of 10 young massive clusters in M82 galaxy from their
corresponding HII regions radii and found $\eta_\mathrm{he} \lesssim 10\%$.
Widths of recombination lines associated with super star clusters in the Antennae
galaxies observed by \citet{2007ApJ...668..168G} also suggest that
$\eta_\mathrm{he} \lesssim 10\%$. \citet{2016ApJ...825..118T} argue that
the heating efficiency is effectively low if majority of massive stars end up
as interacting binaries resulting in much lower stellar wind velocities. On the
other hand, \citet{2009ApJ...697.2030S} estimated the total energy of winds of
young massive clusters in M82 and found rather high values of $\eta_\mathrm{he} = 30 -
100\%$. The second parameter, mass loading $\eta_\mathrm{ml}$, describes an
additional influx of primordial gas into the hot thermalized winds representing
processes as for instance evaporation of dense pre-existing clouds inside the
cluster or evaporation of envelopes and disks of young low-mass protostars and
stars. A similar parameter was introduced in \citet{2010ApJ...711...25S}. Mass
loading is normalized by the mass insertion rate due to stellar winds,
$\dot{M}_\mathrm{SC}$. Both parameters ($\eta_\mathrm{he}$ and
$\eta_\mathrm{ml}$) are not particularly well constrained. Therefore, we let
them vary in a wide range of values and explore the parameter space to
understand how our results depend on them.

Our model predicts the formation of secondary stellar generations within the
cluster, with the matter injected by the winds from massive stars of the
first generation. This may be related to the multiple stellar populations found in
globular clusters \citep[see e.g.][and references therein]{2004ApJ...605L.125B,
2007ApJ...661L..53P, 2016arXiv160609468B}, and recently also in intermediate age
massive clusters \citep[e.g.][]{2009A&A...497..755M}. Additionally,
spectroscopic observations revealed anti-correlations between certain pairs of
light elements \citep[e.g. sodium and oxygen; see][]{2006A&A...450..523C,
2009A&A...505..139C} suggesting that a fraction of stars in globular clusters
could form out of gas enriched by products of high-temperature H-burning, as
those produced either in massive stars \citep{2007A&A...475..859D} or
in Asymptotic Giant Branch stars \citep{2010MNRAS.407..854D}. Our model is not
directly applicable to the formation of globular cluster, because we set the
metallicity of the first generation stars to be solar for the sake of
comparability with our previous works on super star clusters. However, the
qualitative predictions of mass accumulation and secondary star formation are
significant and robust. Furthermore, the rapidly cooling winds model
presents multiple features that may help to eliminate some of the problems
encountered by other self-enrichment scenarios. Specifically, it provides a
mechanism to capture fast ($\gtrsim 1000$\,km\,s$^{-1}$) stellar winds
inside the cluster and it self-consistently predicts that under certain
conditions, secondary star formation occurs in the very center of clusters. On
the other hand, no completely satisfactory explanation of the multiple
populations found recently in globular clusters exists, and this model would
also suffer by similar problems as other scenarios based on self-enrichment by
massive stars \citep[see][and references therein]{2015arXiv151001330B}.

The paper is organized as follows: in \S\ref{sec:physmod} we describe the
adopted physical model of the cluster, \S\ref{sec:semianl} and \S\ref{sec:num}
introduce the semi-analytic and numerical codes used to calculate the model,
respectively. Our results are presented in \S\ref{sec:results}. Specifically, we
give semi-analytic estimates of the mass necessary for self-shielding
(\S\ref{ssec:selfsh}), describe the radiation-hydrodynamic simulations
confirming the estimates and calculate the synthetic emission line spectra of
the simulated clusters (\S\ref{ssec:sim}), use the semi-analytic estimates to
explore a larger parameter space and predict certain properties of secondary
stellar generations (\S\ref{ssec:parspace}). The implications of the results for
the evolution of massive star clusters is discussed in \S\ref{sec:discussion},
and our conclusions are formulated in \S\ref{sec:conclusions}.


\section{Physical model}
\label{sec:physmod}


\begin{deluxetable*}{ccl}
\tablecaption{Parameters of all discussed models.
\label{tab:compar}}
\tablehead{parameter & value & description}
\startdata
$M_\mathrm{SC}$ & $10^7$\,M$_\odot$ & mass of the first stellar generation (1G)\\
$\beta$ & $1.5$ & slope of the 1G stellar density distribution\\
$R_c$ & $1.578$\,pc & core radius of the 1G radial distribution \\
$R_\mathrm{SC}$ & $4.5$\,pc & cluster radius, cutoff of the 1G radial distribution \\
$R_\mathrm{TH}$\tablenotemark{a} & $3.0$\,pc & radius of the cluster with top-hat 1G radial distribution \\
$Z_0$ & 0.02 & metallicity of 1G stars \\
$\mu_\mathrm{i}$ & 0.609 & mean mol. weight of hot and warm gas, $T \ge 10^4$\,K \\
$\mu_\mathrm{c}$ & 2.35 & mean mol. weight of cold gas, $T < 10^4$\,K \\
$\mu_\mathrm{H}$\tablenotemark{b} & 1.273 & mean mol. weight per hydrogen nuclei \\
$\eta_\mathrm{he}$\tablenotemark{c} & $0.001 - 1$ & heating efficiency \\
$\eta_\mathrm{ml}$\tablenotemark{d} & $0 - 5$ & mass loading \\
\enddata
\tablenotetext{a}{$R_\mathrm{TH}$ is chosen so that the half-mass radius is the
same as in the case of cluster with Schuster 1G distribution with parameters
$R_c$ and $R_\mathrm{SC}$ given above.}
\tablenotetext{b}{Used for calculating $n_i$ and $n_e$ in cooling rate $Q$ and
recombination rate $\dot{N}_\mathrm{r}$.}
\tablenotetext{c}{$\eta_\mathrm{he} = 0.05$, $0.1$ and $0.3$ for models A, B and
C, respectively.}
\tablenotetext{d}{$\eta_\mathrm{ml} = 1$ for models A, B and C.}
\end{deluxetable*}

We consider a young cluster with a first stellar generation (1G) of mass
$M_\mathrm{SC} = 10^7$\,M$_\odot$ formed abruptly at time $t=0$. The stellar
density $\rho_\mathrm{\star}$ is given by the spherically symmetric Schuster
distribution in a form
\begin{equation}
\rho_\star = M_\mathrm{SC} f_\mathrm{Sch}(r,\beta,R_{c},R_\mathrm{SC})
\label{eq:stardens}
\end{equation}
\begin{equation}
f_\mathrm{Sch}(r,\beta,R_{c},R_\mathrm{SC}) = \left\{
\begin{array}{lll}
\frac{C_\mathrm{Sch}}{\left[1 + (r/R_c)^2\right]^{\beta}} & \mathrm{for} & r \le R_\mathrm{SC} \\
0 & \mathrm{for} & r > R_\mathrm{SC} \\
\end{array}
\right.
\label{eq:schuster}
\end{equation} 
\begin{equation}
C_\mathrm{Sch} = \frac{3}{4\pi} R_\mathrm{SC}^{-3} 
\left[{}_{2}F_{1}(\frac{3}{2},\beta,\frac{5}{2},-\frac{R_\mathrm{SC}^2}{R_{c}^2})\right]^{-1}
\end{equation}
where ${}_{2}F_{1}$ is the Gauss hypergeometric function,
${}_{2}F_{1}(3/2,\beta,5/2,-(r/{R_{c})^2})$ hereafter abbreviated as
$F_\beta(r)$, and $C_\mathrm{Sch}$ is the normalization constant. It has been
shown by \citet{1998SerAJ.158...15N} that the Schuster distribution with
the slope $\beta = 1.5$ (see Equation \ref{eq:schuster}) approximates well the
King stellar surface density profile \citep{1962AJ.....67..471K}. The King
profile was originaly obtained for the globular clusters, however, it is also in
a good agreement with the observed stellar surface density profiles of young
massive clusters \citep[e.g.][]{2009A&A...501..563E}. Therefore, we use $\beta =
1.5$ for the all presented models. Values of the core radius, $R_c = 1.58$\,pc,
and the cluster radius, $R_\mathrm{SC} = 4.5$\,pc, result in the half mass
radius $R_h = 2.38$\,pc which is the same as for the uniform sphere with
radius $3$\,pc. The corresponding gravitational potential $\Psi_\star$ can be
obtained using a standard formula for the potential of the spherically symmetric
density distribution \citep[see e.g.][]{2008gady.book.....B}
\begin{eqnarray}
\Psi_\star(r) & = & -GM_\mathrm{SC} \left\{
r^2 F_\beta(r) + \frac{3R_c^{2\beta}}{2(\beta-1)}
\left[ (R_c^2 + R_\mathrm{SC}^2)^{1-\beta} 
\right. \right.
\nonumber\\
& - & 
\left. \left.
(R_c^2+r^2)^{1-\beta} \right]
\right\}
\label{eq:psi}
\end{eqnarray}
where $G$ is the gravitational constant. The escape velocity from radius $r$ to
infinity is $v_\mathrm{esc}(r) = \sqrt{|2\Psi_\star(r)|}$, and the free fall time from
radius $r$ to the center is 
\begin{equation}
t_\mathrm{ff}(r) = \int_0^{r} \left(2 [\Psi_\star (r) - \Psi_\star (r')]\right)^{-1/2} dr' \ .
\label{eq:tff}
\end{equation}


Massive stars insert mass and mechanical energy into the cluster volume
through their radiation driven winds. Following \citet{1985Natur.317...44C} we
assume that mutual collisions of individual stellar winds result in a hot gas
filling most of the cluster volume (as validated numerically by
\citealt{2000ApJ...536..896C} and others). We model this process by inserting
mass and  thermal energy smoothly  into the cluster with  total rates
$\dot{M}_\mathrm{SC}$ and $L_\mathrm{SC}$, respectively. We assume that the
sources follow the stellar distribution $\rho_\star$ and hence the mass and
energy insertion rate densities are, respectively
\begin{equation}
q_m(r) = (1 + \eta_\mathrm{ml}) \dot{M}_\mathrm{SC}
f_\mathrm{Sch}(r,\beta,R_{c},R_\mathrm{SC}) \ ,
\label{eq:qm}
\end{equation} 
\begin{equation}
q_e(r) = \eta_\mathrm{he} L_\mathrm{SC}
f_\mathrm{Sch}(r,\beta,R_{c},R_\mathrm{SC}) \ .
\label{eq:qe}
\end{equation}
Here we use two additional parameters described in \S\ref{sec:intro}: the
heating efficiency $\eta_\mathrm{he}$, which defines the fraction of the
mechanical energy of individual stellar winds that is converted into the cluster
wind thermal energy, and the mass loading $\eta_\mathrm{ml}$, which specifies
the amount of additional material complementing the reinserted wind.
In \S\ref{ssec:selfsh} and \S\ref{ssec:sim}, we study in detail three models
with $\eta_\mathrm{ml} = 1$ and $\eta_\mathrm{he} = 0.05$, $0.1$ and $0.3$ (see
Table~\ref{tab:compar}). Furthermore, we discuss the most important features of
the model for a large range of these parameters $\eta_\mathrm{he} \in (0,1)$,
$\eta_\mathrm{ml} \in (0,5)$ in \S\ref{ssec:parspace}.

The total amounts of mass and energy inserted into the cluster per unit time,
$\dot{M}_\mathrm{SC}$ and $L_\mathrm{SC}$, respectively, are functions of time.
They are determined using the stellar population synthesis code Starburst99
\citep{1999ApJS..123....3L} by the procedure described in detail in
\citet{2011ApJ...740...75W}. It is assumed that 1G stars were formed
instantaneously with the standard Kroupa initial mass function
\citep{2001MNRAS.322..231K}; the Geneva stellar evolution tracks with the high
mass loss wind model are used \citep[see][for details]{1992ApJ...401..596L}. The
metallicity of the gas, $Z$, is also time-dependent however, at any given time,
we assume that it presents a uniform value within the whole computational
domain. This is given by equation:
\begin{equation}
Z(t) = \frac{\dot{M}_\mathrm{metals} + \eta_\mathrm{ml}Z_0\dot{M}_\mathrm{SC}}
{(1+\eta_\mathrm{ml})\dot{M}_\mathrm{SC}}
\label{eq:Z}
\end{equation}
where $Z_0 = Z_\odot$ is the metallicity of the first stellar generation and
$\dot{M}_\mathrm{metals}$ is the total amount of elements heavier than He
inserted by massive stars per unit time as provided by Starburst99. Note that it is always 
$\dot{M}_\mathrm{metals} \ge Z_0\dot{M}_\mathrm{SC}$ (metallicity of stellar
winds is at least the metallicity of the gas from which the stars were formed),
and therefore $Z(t) \ge Z_0$ all the time.

The gas inserted into the cluster by 1G stars rapidly establishes a star cluster
wind. This process is described by the hydrodynamic equations including terms for
energy losses due to radiative cooling and eventually for gravity (of
stars and self-gravity of the gas). These equations are accompanied by the ideal
gas equation of state in a form
\begin{equation}
P = \frac{\rho k_\mathrm{B}T}{\mu m_\mathrm{H}} \ ,
\label{eq:eos}
\end{equation}
where $P$, $\rho$ and $T$ are gas pressure, density and temperature,
respectively; $k_\mathrm{B}$ is the Boltzmann constant and $m_\mathrm{H}$ is the
hydrogen nuclei mass. The mean molecular weight $\mu$ is either $\mu = \mu_i =
0.609$ for the hot and warm gas with $T \ge 10^4$\,K (assuming it is ionized) or
$\mu = \mu_c = 2.35$ for smaller temperatures (assuming it is molecular).
We do not consider $\mu$ of the atomic phase, because the semi-analytic model
describes only the hot ionized gas, and because the mass of the atomic phase is
negligible in numerical models as the gas densities are so high that the gas
shielded\footnote{We use the words "shielded", "shielding" and "self-shielding"
in the following way: the gas is called shielded when all ionizing radiation has
been absorbed before reaching it; the gas is called shielding when it is
absorbing the radiation that keeps it warm and ionized. We call the whole object
(usually clump or stream) self-shielding when it consists of both shielding and
shielded gas.} against ionizing radiation cools down to $\sim 10$\,K almost
immediately. We define also the mean molecular weight per hydrogen nuclei
$\mu_\mathrm{H} = 1.273$ and use it to calculate electron and ion particle
densities needed for radiative cooling computations: $n_e = n_i =
\rho/\mu_\mathrm{H}$. 

The exact form of the hydrodynamic equations for the semi-analytic model differs
from those obtained for the numerical model, because the former ones are 1D
spherically symmetric and time independent, while the latter ones are 3D time
dependent and include more physical effects as for instance self-gravity and EUV
radiation. Both sets of equations are explicitly given in sections
\S\ref{sec:semianl} and \S\ref{sec:num}, respectively. All discussed
parameters of the model are summarized in Table~\ref{tab:compar}.

\begin{figure}
\plotone{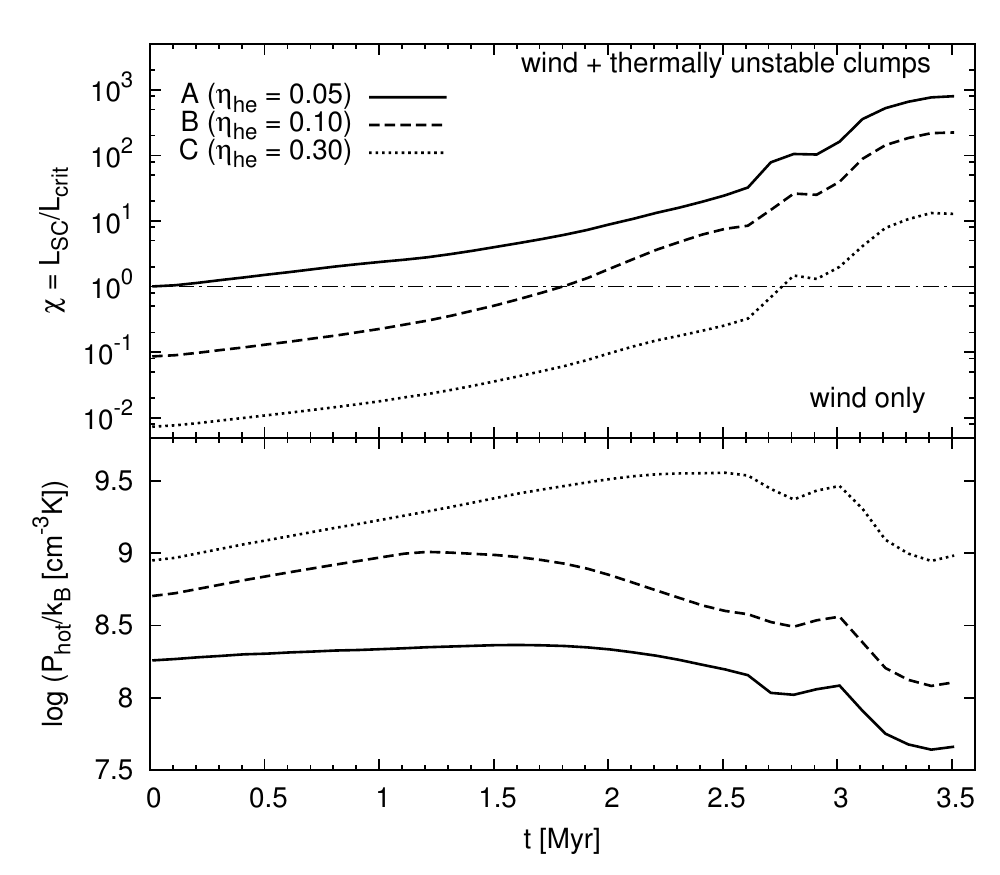}
\caption{Top: evolution of the ratio of the cluster mechanical luminosity to the
critical luminosity, $\chi \equiv L_\mathrm{SC}/L_\mathrm{crit}$, for the
models A (solid), B (dashed) and C (dotted) with parameters given in
Table~\ref{tab:compar}. The dash-dotted line is the $\chi
= 1$ line separating regions with the quasi-adiabatic behavior (below) and
the rapidly cooling winds (above). Bottom: evolution of the mean pressure of the hot
gas below $R_\mathrm{SC}$ approximated by the pressure at the stagnation radius
calculated by the semi-analytic code for a cluster with top-hat stellar density
profile and $R_\mathrm{TH} = 3$\,pc.
}
\label{fig:xiPhot}
\end{figure}


\section{Semi-analytic code}
\label{sec:semianl}


We use the semi-analytic code developed by \citet{2004ApJ...610..226S} to
estimate the amount of mass accumulated inside the cluster. This code
searches for the stationary solution for the hot star cluster wind and in case
it does not exist, it calculates how much mass has to be removed from the hot
phase to allow the stationary solution existence. It is computationally much
cheaper than full radiation-hydrodynamic simulations and therefore it allows us
to explore the parameter space in \S\ref{ssec:parspace}. The procedure to use
the code for considering an evolving cluster is described by
\citet{2011ApJ...740...75W} and we briefly summarize it here for the convenience
of the reader. The implementation of the semi-analytic code allows
only sources with a top-hat radial density profiles, i.e. the mass and energy
deposition rate densities, $q_m$ and $q_e$, are spatially constant within the
cluster. Therefore, clusters with the Schuster density profile used in this work
are approximated by top-hat density profiles with the same half-mass radius $R_h
= 2.38$\,pc resulting in the top-hat cluster radius $R_\mathrm{TH} = 3$\,pc. The
gravity acting on the hot gas is neglected, since the thermal energy of the
hot gas is always higher than its potential energy in the gravitational field of
the cluster. However, the gravity is taken into account in estimates of the
fraction of the reinserted mass that stays in the cluster (see Equations
(\ref{eq:resc}) and (\ref{eq:macc}) below).

We assume that mass and energy deposition rate densities, $q_m$ and $q_e$, vary
on a substantially longer time-scale than the cluster wind crossing time. Then,
for a cluster at a given time, we can search for a stationary wind solution by
solving the spherically symmetric hydrodynamic equations:
\begin{equation}
\frac{1}{r^2}\frac{d}{dr}(\rho u r^2) = q_m
\label{eq:basic_con}
\end{equation}

\begin{equation}
\rho u \frac{du}{dr} = - \frac{dP}{dr} - q_m u
\label{eq:basic_mom}
\end{equation}

\begin{equation}
\frac{1}{r^2} \frac{d}{dr}\left[
\rho u r^2 \left( \frac{u^2}{2} + \frac{\gamma}{\gamma-1}\frac{P}{\rho} \right)
\right] = q_e - Q
\label{eq:basic_ener}
\end{equation}
where $\rho$, $u$ and $P$ are the wind density, velocity and pressure, respectively.
The energy equation (\ref{eq:basic_ener}) includes the cooling term $Q = n_i n_e
\Lambda(T, Z)$ where $n_i = n_e = \rho/\mu_\mathrm{H}$ are the ion and electron number
densities, $Z$ is the gas metallicity given by Eq.~\ref{eq:Z} and $\Lambda(T,Z)$
is a cooling function calculated by \citet{2009A&A...508..751S}. The pressure is
calculated using the ideal gas equation of state (\ref{eq:eos}) with $\mu_i =
0.609$.

As shown by \citet{2007ApJ...658.1196T}, Equations~(\ref{eq:basic_con}) --
(\ref{eq:basic_ener}) have a solution for all radii only if the cluster
mechanical luminosity, $L_\mathrm{SC}$, does not exceed a certain critical
value, $L_\mathrm{crit}$. This critical luminosity can be found using a
bisection method by varying $L_\mathrm{SC}$ and checking whether the solution of
(\ref{eq:basic_con}) -- (\ref{eq:basic_ener}) exists or not. In order to obtain
a more accurate value of $L_\mathrm{crit}$ for clusters with a Schuster density
profiles, we corrected $L_\mathrm{crit}$ by a constant factor $3.5$ found by
comparison with 1D numerical simulations by \citet{2013ApJ...772..128P} (see
their Figure~6). We define a ratio between the cluster mechanical luminosity and
its critical value, $\chi = L_\mathrm{SC} / L_\mathrm{crit}$, and plot its time
evolution for models A, B and C in Figure~\ref{fig:xiPhot} (top panel).

If $\chi > 1$, the stationary solution of Equations (\ref{eq:basic_con}) --
(\ref{eq:basic_ener}) does not exists for the whole cluster volume.
However, it is still possible to find a solution in the region $r >
R_\mathrm{st}$ where $R_\mathrm{st}$ is the so called stagnation radius below
which the wind velocity is zero. In \citet{2011ApJ...740...75W} we assumed that
all gas inserted below $R_\mathrm{st}$ accumulates inside the cluster. 
Here we find through hydrodynamic simulations (see \S\ref{ssec:sim} below) that
models with a negative stellar radial density gradient (e.g. Schuster profiles)
behave differently. The wind velocity is positive in the whole cluster volume
and no stagnation radius exists even for $\chi > 1$. In our calculations,
dense clumps are still formed through thermal instabilities inside the cluster.
As they have positive radial velocities "inherited" from the wind gas from
which they form, in the absence of gravity, they would leave the cluster.
However, if gravity is taken into account, clumps  formed at smaller radii, with
radial velocities smaller than the escape velocity (see below), are captured and
fall into the cluster center.

The fraction of the inserted gas that ends up in dense clumps can be estimated
by comparing the mass deposition rate density of a given model $q_m$ with the
corresponding quantity $q_{m,\mathrm{crit}}$ of the model with the same
parameters but a mechanical luminosity $L_\mathrm{SC}$ equal to
$L_\mathrm{crit}$. This is because clump formation effectively lowers the
density of the hot medium down  to the  level obtained when $\chi = 1$. The
clumps acquire positive radial velocities similar to that of the wind at the
radius where they form. Thus we define the escape radius $R_\mathrm{esc}$ as the
distance from the cluster center where the wind velocity $u$  equals  the
cluster escape velocity:
\begin{equation}
u(R_\mathrm{esc}) = v_\mathrm{esc}(R_\mathrm{esc}) \equiv \sqrt{|2\Psi_\star(R_\mathrm{esc})|}
\label{eq:resc}
\end{equation}
where $\Psi_\star$ is given by Equation~(\ref{eq:psi}). Then, we assume that all the
clumps that form below $R_\mathrm{esc}$ are captured and clumps that form above
$R_\mathrm{esc}$ leave the cluster with the wind. Therefore, the amount
of gas accumulated up to a certain time $t$ is estimated as 
\begin{equation}
M_\mathrm{acc}(t) = \int_{t_{bs}}^t\int_0^{R_\mathrm{esc}} [q_m(r, t') - q_{m,crit}(r, t')] dr dt'
\label{eq:macc}
\end{equation}
where $t_\mathrm{bs}$ is the time at which $\chi$ exceeds $1$ and thermal
instabilities start to appear inside the cluster. Note that in this
approach, the accumulated mass, $M_\mathrm{acc}$, is overestimated as it
ignores the hydrodynamic forces from the wind pushing the clumps outwards. On
the other hand, $M_\mathrm{acc}$ is underestimated as Equation~(\ref{eq:macc})
ignores the gravitational force caused by the accumulated gas and the forming
secondary stellar generation. However, a comparison for models A, B and C,
between $M_\mathrm{acc}$ and $M_\mathrm{acc}^\mathrm{num}$ obtained from
numerical simulations that include both the above effects, suggests that the
errors are not large (see Table~\ref{tab:sims} and Figure~\ref{fig:mevol}).


\section{Numerical code}
\label{sec:num}


The numerical model is based on the three-dimensional, adaptive mesh refinement
(AMR) code FLASH v4.2.1 \citep{2000ApJS..131..273F}. The AMR is handled by the
PARAMESH library \citep{2000CoPhC.126..330M}, the whole code is parallelized via
domain decomposition under the Message Passing Interface (MPI). The hydrodynamic
equations are solved using a modified version of the Piecewise Parabolic Method
\citep[PPM][]{1984JCoPh..54..174C} with the time-step controlled by the
Courant-Friedrichs-Lewy criterion. They have a form
\begin{equation}
\frac{\partial \rho}{\partial t} + \nabla\cdot \rho\mathbf{u} = q_m
\label{eq:flash_con}
\end{equation}
\begin{equation}
\frac{\partial \rho\mathbf{u}}{\partial t} + \nabla\cdot\rho\mathbf{uu} + \nabla
P = \rho\nabla\Psi - q_m \mathbf{u}
\label{eq:flash_mom}
\end{equation}
\begin{equation}
\frac{\partial \rho E}{\partial t} + \nabla\cdot(\rho E + P)\mathbf{u} =
\rho\mathbf{u}\cdot\mathbf{g} + q_e - Q
\label{eq:flash_ener}
\end{equation}
where $\rho$, $\mathbf{u}$ and $P$ are the gas density, velocity and pressure,
respectively, and $E = P/[(\gamma-1)\rho] + u^2/2$ is the total energy per unit
mass with $\gamma$ being the ratio of specific heats. The mass and energy
deposition rate $q_m$ and $q_e$ are given by Equations~(\ref{eq:qm}) and
\ref{eq:qe}, respectively, and their time evolution is obtained from Starburst99
code as described in \S\ref{sec:semianl}. The cooling term $Q$ is
calculated using a procedure based on sub-cycling described in
\citet{2008ApJ...683..683W}. The gravitational potential $\Psi$ consists of
three parts, $\Psi = \Psi_\star + \Psi_\mathrm{s} + \Psi_\mathrm{g}$, where
$\Psi_\star$ is the potential of 1G stars given by Equation~(\ref{eq:psi}),
$\Psi_\mathrm{s}$ is the potential of sink particles (see below) calculated by
direct force summation, and $\Psi_\mathrm{g}$ is the potential of the gas
obtained by solving the Poisson equation
\begin{equation}
\nabla^2 \Psi_\mathrm{g} = 4\pi G\rho \ .
\label{eq:poisson}
\end{equation}
Equation~(\ref{eq:poisson}) is solved using the tree code algorithm described in
W\"unsch et al. (2017, in prep.), it also provides the gravitational
acceleration $g$ corresponding to $\Psi$. The set of
Equations~(\ref{eq:flash_con})--(\ref{eq:flash_ener}) is closed by the equation
of state as in (\ref{eq:eos}) with mean molecular weights $\mu_i$ and $\mu_c$
for the appropriate temperature regimes (see Table~\ref{tab:compar}).

The ionizing radiation of stars is included using module OpticalDepth of the
radiation transport code TreeRay described in W\"unsch et al. (2016, in prep.).
Instead of calculating the radiation transport exactly, it assumes that the
whole computational domain is embedded in a bath of ionizing radiation with a
uniform photon flux coming from all directions. The photon flux is approximated
by flux $F_\mathrm{UV}$ in the center of a sphere with radius $R_\mathrm{TH}$
and uniform distribution of radiation sources with the total photon production
rate $\dot{N}_\mathrm{UV}$ given by the Starburst99 code for clusters with
the selected parameters (see also Equation (19) in \citealt{2014ApJ...792..105P})
\begin{equation}
F_\mathrm{UV} = (3 \dot{N}_\mathrm{UV})/(16\pi R_\mathrm{TH}^3) \ .
\end{equation}
Using the generalized algorithm TreeCol developed by \citet{2012MNRAS.420..745C},
the OpticalDepth module calculates the emission measure $EM_j$ for each
grid cell and for each direction
\begin{equation}
EM_{j} = \int_0^\infty \left(\frac{\rho(s,j)}{\mu_\mathrm{H}}\right)^2 ds
\end{equation}
where index $j$ runs over 12 directions, the minimum number defined by the
HealPix library \citep{2005ApJ...622..759G} and $\rho(s,j)$ is the gas density
in direction $j$ at distance $s$ from the cell. Invoking the on-the-spot
approximation \citep{1974agn..book.....O} and assuming that EUV photons are
destroyed along the incoming ray by recombinations to other than the fundamental
level, the number of ionizing photons $n_\mathrm{UV}$ entering the grid cell is
\begin{equation}
n_\mathrm{UV} = \sum_{j=1}^{12} A_\mathrm{surf,j} H(F_\mathrm{UV} - EM_j/\alpha_\mathrm{B})
\end{equation}
where $A_\mathrm{surf,j} = [(\pi/48)^{(1/2)}\mathrm{d}V]^{(2/3)} $ is a fraction of the
grid cell surface associated with direction $j$ with $\mathrm{d}V$ being the
grid cell volume, $\alpha_\mathrm{B} = 2.7\times 10^{-13}$\,cm$^{3}$s$^{-1}$ is
the recombination coefficient into excited states only and $H$ is the Heaviside
step function. Subsequently, the grid cell is assumed to be ionized and its
temperature is maintained at $T_i = 10^4$\,K if $n_\mathrm{UV} > 0$. Otherwise,
if $n_\mathrm{UV} = 0$, the grid cell is allowed to cool to lower temperatures
(which in the majority of cases means that it quickly cools down to the minimum
allowed temperature $10$\,K because of its high density).

As the cold gas evolves under the influence of its own gravity it may become
gravitationally unstable and collapse if its mass exceeds the Jeans mass.
Therefore we include the sink particles module of FLASH
\citep{2010ApJ...713..269F}. If the gas density in a certain grid cell exceeds a
threshold $\rho_\mathrm{sink}$ and if the gas within the so called accretion
radius, $r_\mathrm{acc}$, fulfills a number of criteria, a sink particle is
created. The criteria are: (i) the cell is at the highest refinement level, (ii)
the cell represents a local minimum of $\Psi_g$, (iii) the mass exceeds the
Jeans mass, (iv) the flow is converging ($\nabla\cdot\mathbf{u}<0$), (v) the gas
is gravitationally bound, and (vi) the region does not overlap with some other
sink particle. Additionally, a fraction of gas with density exceeding
$\rho_\mathrm{sink}$ within $r_\mathrm{acc}$ of each particle is accreted onto
it, i.e. the gas density is truncated to $\rho_\mathrm{sink}$ and the mass is
added to the mass of the sink particle. For all models presented here we set
$\rho_\mathrm{sink} = 10^{-17}$\,g\,cm$^{-3}$ and $r_\mathrm{acc} = 0.05$\,pc
corresponding to $2.5\times$ grid cell size, as recommended by authors of the
sink particles module of the code. With these values, it is not  possible to
follow the fragmentation process down to the mass of individual stars.
Therefore, sink particles here represent "clusters" or "associations" of
secondary stellar generations rather than individual stars. The number of sink
particles depends on the simulation resolution, however, the total mass in sink
particles does not. We checked this by comparing runs A, B and C with their
low-resolution counterparts calculated for the whole time at $128^3$ grid. 

We simulate models A, B and C for their first $3.5$\,Myr of evolution, \ie
before SNe start to explode. The computational domain has size ($10$\,pc)$^3$
and all outer boundary conditions are set to outflow. Most of the time is
computed on a uniform $128^3$ grid, however, several selected periods are
calculated with AMR at the maximum resolution corresponding to $512^3$.


\subsection{Synthetic spectra}
\label{ssec:spectra}


We calculate the synthetic spectra of a hydrogen recombination line formed in
the warm gas (mainly in thermally unstable clumps) present in the
simulations. This will allow to compare the calculated  models with radio
observations (e.g. in mm-wavelengths by ALMA). We choose the H30$\alpha$ with
rest frequency $\nu_0 = 231.9$\,GHz due to its proximity to the frequency of
molecular CO(2-1) line allowing eventually to probe both ionized and molecular
gas with a single observation. We assume that the emission is optically thin and
verify that such is the case afterwards by calculating the maximum optical depth
in the simulations. This allows us to treat each grid cell separately and
calculate its line emission $T_\mathrm{L,cell}$ as \citep{2004tra..book.....R}
\begin{equation}
T_\mathrm{L,cell} = 1.92\times 10^3 \left( \frac{T_e}{K} \right)^{-3/2}
\left( \frac{EM_\mathrm{cell}}{\mathrm{cm}^{-6} \mathrm{pc}} \right)
\left( \frac{\Delta\nu}{\mathrm{kHz}} \right)^{-1}
\label{eq:TL}
\end{equation}
where $EM_\mathrm{cell} = (\rho/\mu_\mathrm{H}/m_\mathrm{H})^2 \mathrm{d}s$ is
the emission measure of the cell, $T_e$ is the electron temperature assumed to
be the same as the gas temperature $T$ in the grid cell, $\mathrm{d}s$ is the
linear cell size and $\Delta\nu = \Delta v\nu_0/c$ is the width of a frequency
bin corresponding to the velocity bin width $\Delta v$. In all calculations
presented here we cover the velocity range $(-400,400)$\,km\,s$^{-1}$ with $500$
bins leading to $\Delta v = 1.6$\,km\,s$^{-1}$ and $\Delta\nu = 1240$\,kHz.

For a given line-of-sight aligned with one of the Cartesian axes, we calculate
the line profile $T_\mathrm{L}$ by summing up all contributions
$T_\mathrm{L,cell}\mathrm{d}s$ intersecting with the line-of-sight and
distributing them into velocity bins according to the velocity of the grid cell
convolved with the Gaussian of width $\sigma = k_\mathrm{B}T/(\mu_e
m_\mathrm{H})$ to account for thermal broadening. We present
position-velocity diagrams of our simulations at $z=0$ plane integrated along
the $y$-direction in Figure~\ref{fig:sim:he2}. Additionally, we sum up
contributions of all lines-of-sight $T_\mathrm{L}(\mathrm{d}s)^2$ and normalize
the result by $\pi D_\mathrm{res}^2/4$ in order to obtain the brightness
temperature profile $T_b$ as seen by a telescope with angular resolution
$D_\mathrm{res}$.

Finally, we calculate the maximum optical depth by summing up individual grid
cell contributions
\begin{equation}
\tau_\mathrm{L,cell} = 1.92\times 10^3 \left( \frac{T_e}{K} \right)^{-5/2}
\left( \frac{EM_\mathrm{cell}}{\mathrm{cm}^{-6} \mathrm{pc}} \right)
\left( \frac{\Delta\nu}{\mathrm{kHz}} \right)^{-1}
\end{equation}
along each line-of-sight for each frame of each simulation. We found that
the integrated value $\tau_\mathrm{L}$ never exceeds $10^{-2}$ justifying the
assumption of an optically thin approximation.


\section{Results}
\label{sec:results}

\subsection{Estimate of the shielding mass}
\label{ssec:selfsh}


The evolution of a growing dense clump immersed in the radiation field of the
cluster was discussed in \citet{2014ApJ...792..105P}. It was shown there that
the clump mass $M_\mathrm{c}$  becomes larger than the shielding mass
$M_\mathrm{c,sh}$, rather early in the cluster evolution. $M_\mathrm{c,sh}$  is
the mass of the clump needed to self-shield its interior
against the ionizing radiation. We concluded that clumps may become quickly
seeds of secondary stellar generations formed out of stellar wind matter
carrying the hydrogen burning products originating in the stellar interiors.

By analyzing simulations of models A, B and C we found that self-shielding of
the dense gas typically appears in two qualitatively different configurations.
One possibility is that the warm dense gas falls into the cluster center,
accumulates there and only when its mass exceeds a certain value, it begins to
shield itself against the EUV radiation, allowing its central regions to cool
down and collapses into sink particles. We call this configuration {\em
self-shielding of the central clump.} Another possibility is that the warm dense
gas infalling towards the center along radial streams becomes self-shielding even
before reaching the center. We call this configuration {\em self-shielding of
infalling streams}. The difference between the two scenarios is astrophysically
interesting, because in the former case, the second stellar generation is formed
only in the very center of the cluster, while in the latter one, the stars of
the second generation are formed everywhere in the 1G cluster volume.

Before we derive an equation for the shielding masses in the two previously
described possibilities, we consider a simple configuration-independent threshold,
$M_\mathrm{max,sh}$ for the maximum mass before the self-shielding occurs, based on the
number of available EUV photons from all stars in the cluster. A more massive
object or a group of objects, must include self-shielding regions regardless the
geometry, as there are not sufficient EUV photons to keep them fully
ionized. Therefore, in the calculations, we assume that a central clump or an
infalling stream becomes self-shielding whenever its mass exceeds
$M_\mathrm{max,sh}$, even in cases when its mass is below geometry dependent
criteria $M_\mathrm{c,sh}$ or $M_\mathrm{s,sh}$ (see below)\footnote{Formally, it could
happen e.g. in the case when the size of the central clump exceeds
$R_\mathrm{SC}$ since the number of EUV photons entering the clump scales with
the clump surface.}. Assuming that the warm gas is in pressure
equilibrium with the hot gas, one can calculate the warm gas mass,
$M_\mathrm{max,sh}$, whose total recombination rate is equal to the total photon
production rate of the cluster
$\dot{N}_\mathrm{UV}$:
\begin{equation}
M_\mathrm{max,sh} =  \frac{\mu_i m_H}{\alpha_\mathrm{B}} 
\frac{k T_4}{P_\mathrm{hot}} \dot{N}_{UV} \ .
\label{eq:Mmaxsh}
\end{equation}
where $T_4 = 10^4$\,K is the temperature of the warm ionized gas, and
$P_\mathrm{hot}$ is the pressure of the hot gas that can be approximated by the
pressure at the stagnation radius calculated by the semi-analytic code. Since
$P_\mathrm{hot}$ regulates the volume and therefore the density of the warm gas
and through it  the total number of recombinations, it is then the quantity that
mainly determines the shielding mass in any configuration.
Figure~\ref{fig:xiPhot} (bottom panel) shows the time evolution of
$P_\mathrm{hot}$.


\subsubsection{Self-shielding of the central clump}
\label{ssec:ssclump}

\begin{figure}
\plotone{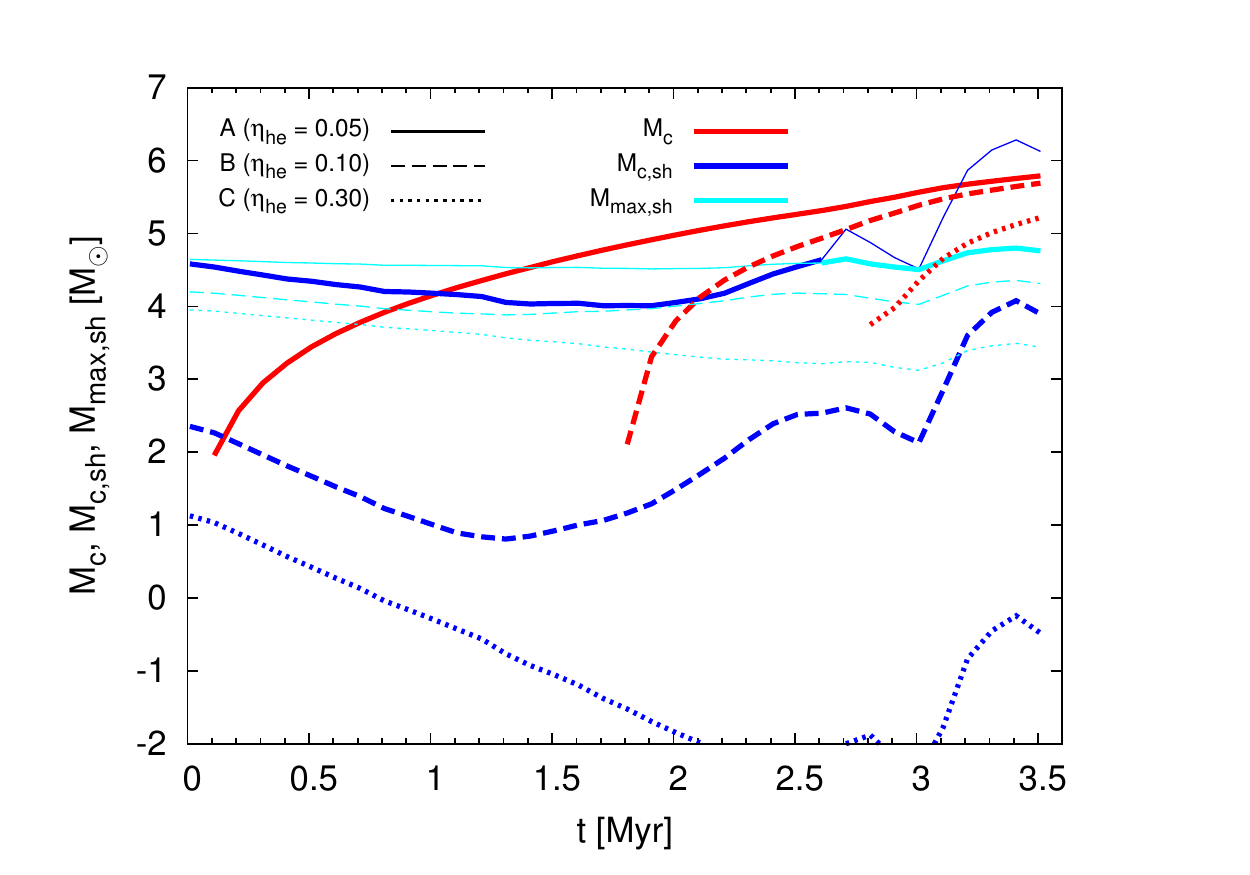}
\caption{Evolution of the mass of the central clump, $M_\mathrm{c}$ (red
curves), and the corresponding shielding mass, which is the smaller (marked
thick) from either $M_\mathrm{c,sh}$ (blue curves) or $M_\mathrm{max,sh}$
(cyan curves). The line types denote models A (solid), B (dashed) and C
(dotted) with parameters given in Table~\ref{tab:compar}. The clump includes a
shielded cold core if $M_\mathrm{c} >
\min(M_\mathrm{c,sh},M_\mathrm{max,sh})$, i.e. if the red curve is above the
blue/cyan curve of the same line type.}
\label{fig:Mc}
\end{figure}

We assume that the mass of the central clump, $M_\mathrm{c}$, is the same as the
total amount of the accumulated gas, $M_\mathrm{acc}$, given by
Equation~(\ref{eq:macc}). Pressure equilibrium between the central clump and hot
gas yields the clump radius
\begin{equation}
R_{c} = \frac{3}{4 \pi} M_\mathrm{c}^{1/3}\left(\frac{k_\mathrm{B} T_4}{\mu_i
m_\mathrm{H} P_\mathrm{hot}}\right)^{1/3} \ .
\label{eq:rc}
\end{equation} 
By comparing the number of EUV photons reaching the clump surface
per unit time with the recombination rate within the whole clump, one 
obtains the central clump shielding mass
\begin{equation}
M_{c,sh} =  \frac{9}{16}\pi \mu_i m_\mathrm{H}
\alpha_\mathrm{B}^{-3} \left(\frac{k_\mathrm{B} T_4}{P_\mathrm{hot}}\right)^5  q_{UV}^3
R^3_{SC}.
\label{eq:mcsh}
\end{equation}

The first $3.5$\,Myr of evolution of $M_\mathrm{c}$ and $M_\mathrm{c,sh}$
for models A, B and C are shown in Fig \ref{fig:Mc}. Note that for
model A which has $\chi > 1$ from the very beginning, the central clump becomes
self-shielding at about $1$\,Myr. For models B and C, the central clump becomes
self-shielding as soon as $\chi$ becomes larger than $1$ at $1.8$ and
$2.8$\,Myr, respectively. This is in a relatively good agreement with the numerical
simulations of these models (see \S\ref{ssec:sim}), even though
self-shielding in model A occurs earlier in the simulation.


\subsubsection{Self-shielding of streams}
\label{ssec:ssstream}


\begin{figure}
\plotone{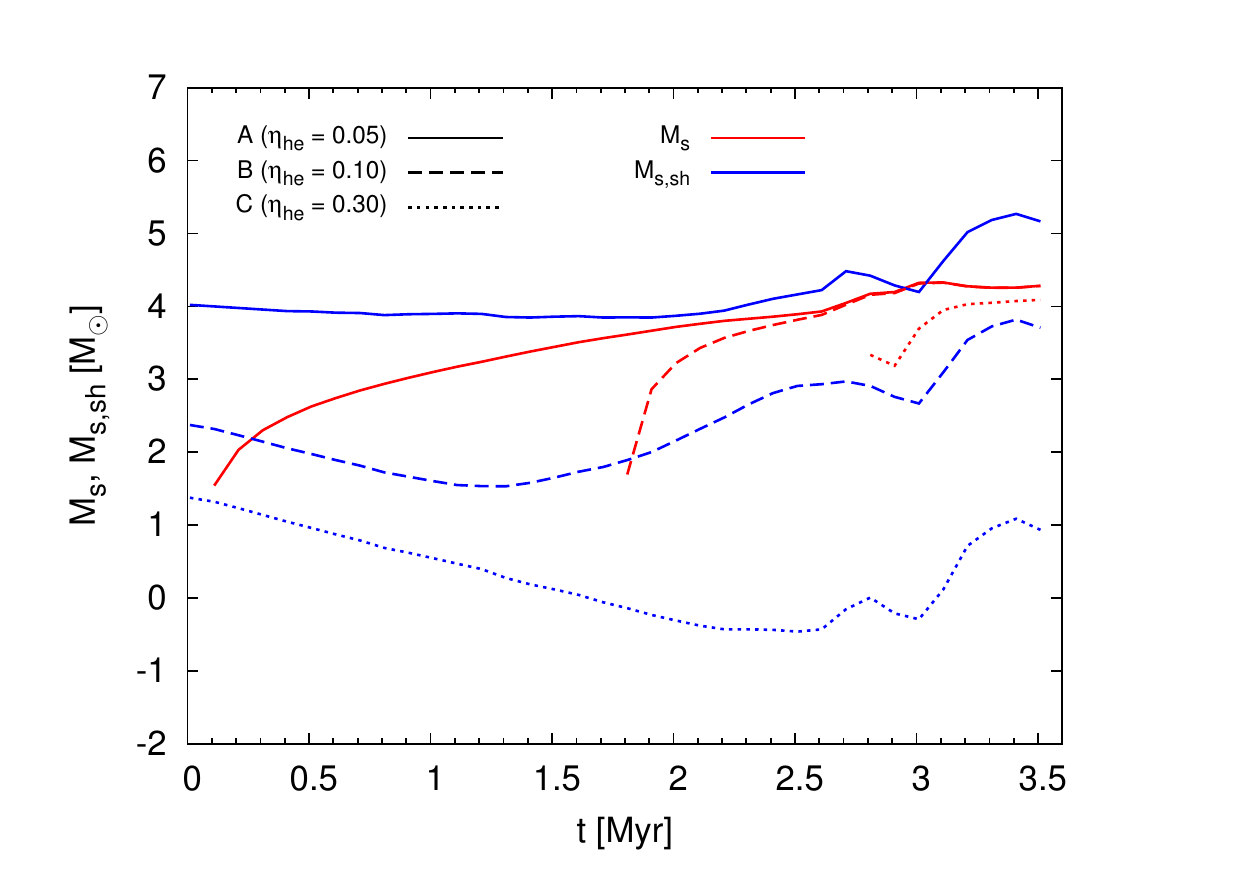}
\caption{Evolution of the mass of the stream, $M_\mathrm{s}$ (red curves), and
the corresponding shielding mass, $M_\mathrm{s,sh}$ (blue curves) for
models A (solid), B (dashed) and C (dotted) with parameters given by
Table~\ref{tab:compar}. The stream includes shielded cold interior if
$M_\mathrm{s} > \min(M_\mathrm{s,sh},M_\mathrm{max,sh})$, i.e. if the red curve
is above the blue curve of the same line type. Note that for the shown models,
it is always $M_\mathrm{s,sh} < M_\mathrm{max,sh}$, and therefore
$M_\mathrm{max,sh}$ is not plotted.}
\label{fig:Ms}
\end{figure}

Since the number of streams of warm gas infalling into the cluster center from
different directions cannot be easily determined, we assume that there is only
one stream into which all gas inserted below the escape radius $R_\mathrm{esc}$
accumulates. This implies  that the calculated stream shielding mass,
$M_\mathrm{s,sh}$, is a lower limit, as more gas is needed
for self-shielding a larger number of streams. 

The amount of gas in the stream, $M_\mathrm{s}$ can be derived by assuming that
the time taken for  the gas to falls into the cluster center is similar to the
free fall time $t_\mathrm{ff}(R_\mathrm{esc})$ given by Equation~\ref{eq:tff}. The
mass of the stream is
\begin{equation}
M_\mathrm{s}(t) = \int_{t-t_\mathrm{ff}}^{t}\int_0^{R_\mathrm{esc}} [q_m(r, t') - q_{m,crit}(r, t')] dr dt',
\label{eq:ms}
\end{equation}
which applies for $t - t_\mathrm{ff} > t_{bs}$ when the hot medium inside the
cluster is thermally unstable.

The ionizing EUV photons from the cluster invade the stream upon reaching its surface, which
is $\pi d R_\mathrm{esc}$, where $d$ is the stream diameter. The number of photons
arriving per unit time and per unit area was estimated by
\citet{2014ApJ...792..105P} as $\frac{1}{4} q_{UV} R_{SC}$, where $q_{UV}$ is
the total EUV photon production rate density of the cluster. At the self-shielding time
$t_\mathrm{s,sh}$, the number of recombinations inside the stream is just in balance
with the total number of incoming ionizing photons
\begin{equation}
\pi \frac{d^2}{4} R_\mathrm{esc} n_{4}^2 \alpha_\mathrm{B} = \pi d
R_\mathrm{esc} \frac{1}{4} q_\mathrm{UV} R_\mathrm{SC},
\label{eq:stream-ballance}
\end{equation}
where $n_4$ is the particle density in the stream computed from the pressure
balance between the stream and the surrounding hot gas $n_4 =
\frac{P_\mathrm{hot}}{k_\mathrm{B} T_4}$. The stream shielding mass is 
\begin{equation}
M_\mathrm{s,sh}(t) = \frac{\pi}{4} d^2 R_{esc} \rho
\label{eq:stream-shield}
\end{equation} 
with the density $\rho = \mu_\mathrm{i} m_\mathrm{H} n_4$. Inserting $d$ from 
Equation~(\ref{eq:stream-ballance}) into Equation~(\ref{eq:stream-shield}) we get
\begin{equation}
M_\mathrm{s,sh}(t) = \frac{\pi}{4} q_\mathrm{UV}^2 R_\mathrm{SC}^2 R_\mathrm{esc}
\alpha_\mathrm{B}^{-2} \mu_i m_\mathrm{H} \left( 
\frac{k_\mathrm{B} T_4}{P_\mathrm{hot}} \right)^3.
\label{eq:mssh}
\end{equation}

The evolution of the $M_\mathrm{s}(t)$ during the first $3.5$\,Myr of the cluster
evolution is compared to the $M_\mathrm{s,sh}$ evolution in Figure~\ref{fig:Ms},
where models A, B and C are shown. For model A, $M_\mathrm{s} $ is always
smaller than $M_\mathrm{s,sh}$ (apart from a very short interval with $M_\mathrm{s}
\lesssim M_\mathrm{s,sh}$ at $3$\,Myr), and thus  the stream is not able to
self-shield, and remains fully ionized throughout the evolution. For models B
and C, the stream mass $M_\mathrm{s}$ is always larger than $M_\mathrm{s,sh}$,
and thus the streams are able to self-shield their interiors immediately after
the start  of thermal instabilities. This is in good agreement with the
numerical simulation in \S\ref{ssec:sim}.


\subsection{Radiation-hydrodynamic simulations}
\label{ssec:sim}


\begin{figure*}
\plotone{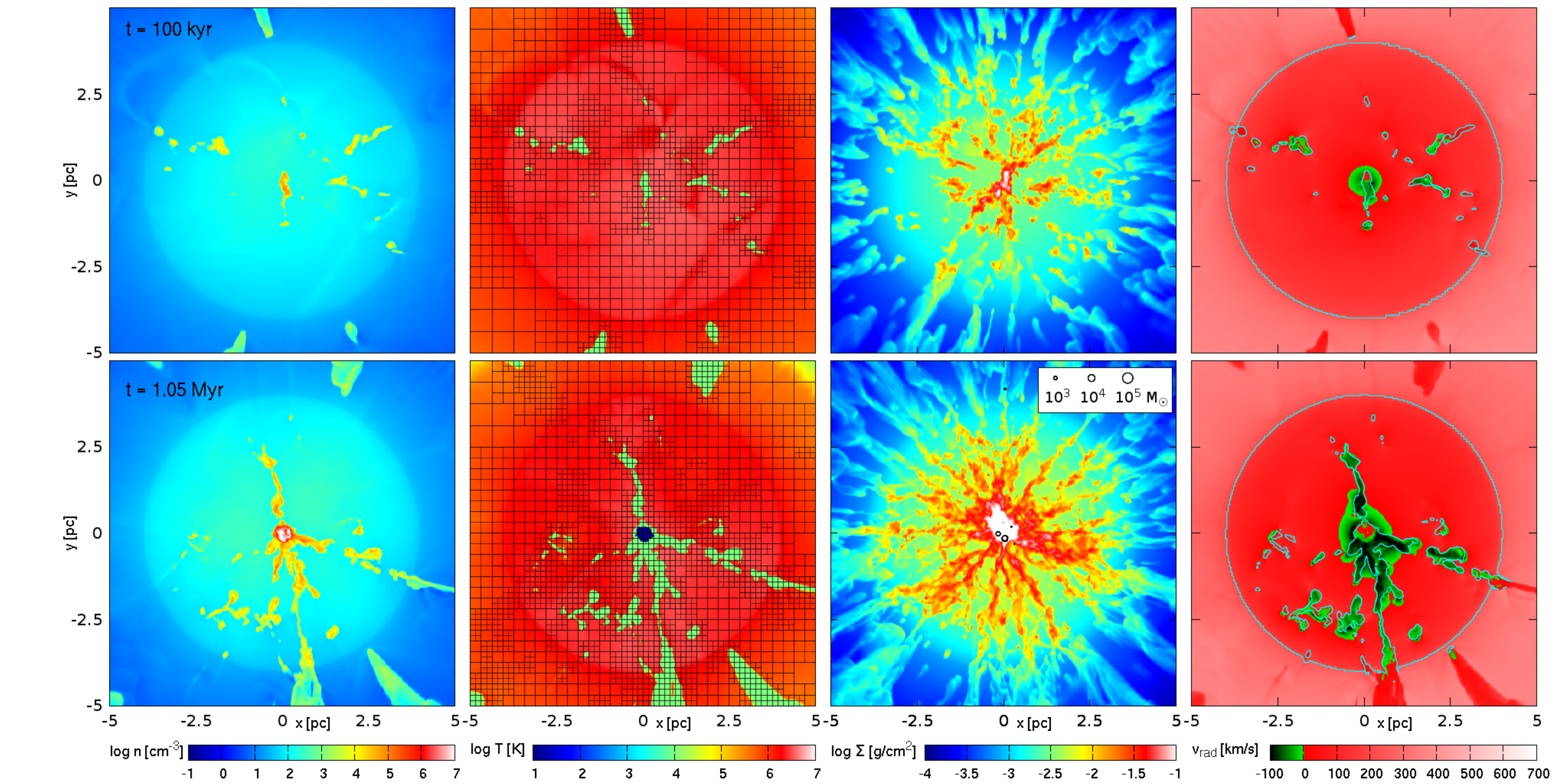}
\caption{RHD simulation of model A shown for times $100$\,kyr (top row of
panels) and $1.05$\,Myr (bottom row). Individual columns of panels show (from left
to right): (i) logarithm of the gas particle density in plane $z=0$, (ii)
logarithm of the gas temperature in the same plane, grid of black lines shows
Flash AMR blocks, (iii) gas column density integrated along the z-direction, and
(iv) the gas radial velocity relative to the cluster center, green-black
corresponds to inwards velocity, red-white is the outwards velocity, the cyan
line separates regions with subsonic and supersonic velocity. The bottom
panel (iii) with the gas column density shows also sink particles represented by
white circles with black borders, and sizes representing their masses as given in
the legend.}
\label{fig:sim:time} 
\end{figure*}

\begin{deluxetable*}{llllllll}
\tablecaption{Properties of studied models.\label{tab:sims}}
\tablehead{
Model & $t_\mathrm{bs}$ & $t_\mathrm{c,sh}$ & $t_\mathrm{s,sh}$ &
$M_\mathrm{tot}$ & $M_\mathrm{acc}$ &
$M_\mathrm{acc}^\mathrm{num}$ &
$M_\mathrm{acc,gas}^\mathrm{num}$ \\
 & (Myr) & (Myr) & (Myr) & ($10^5$\,M$_\odot$) 
 & ($10^5$\,M$_\odot$) & ($10^5$\,M$_\odot$) 
 & ($10^5$\,M$_\odot$) \\
(1) & (2) & (3) & (4) & (5) & (6) & (7) & (8)
}
\startdata
A & --  & 1.0 & --  & 6.9 & 6.2 & 6.1 & 0.14 \\
B & 1.8 & 1.8 & 1.8 & 6.9 & 4.8 & 4.7 & 0.18 \\
C & 2.8 & 2.8 & 2.8 & 6.9 & 1.3 & 1.2 & 0.24 \\
\enddata
\tablecomments{
Columns: (1) Model name. (2) Time of the beginning of the thermal instability,
$t_\mathrm{bs}$. (3) Beginning of the central clumps self-shielding,
$t_\mathrm{c,sh}$. (4) Beginning of the infalling stream self-shielding,
$t_\mathrm{s,sh}$. (5) Total amount of reinserted gas including mass loading.
(6) Amount of the accumulated gas (semi-analytic model). (7) Amount of the
accumulated gas (numerical model). (8) Amount of the remaining warm gas in the
computational domain at the end of the evolution (numerical model).
}
\end{deluxetable*}

\begin{figure*}
\includegraphics[width=1.0\textwidth]{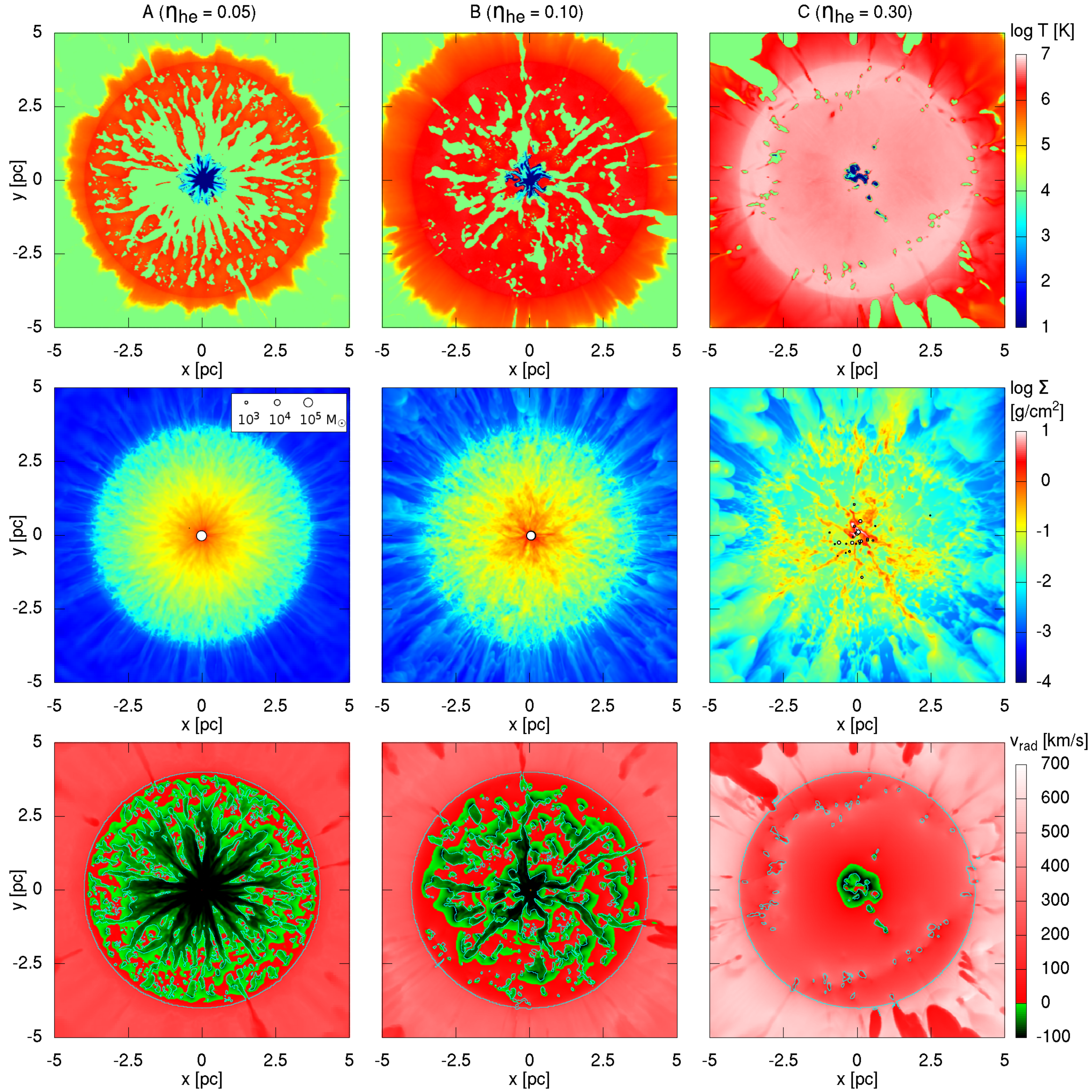}
\caption{Comparison of RHD simulations of models A, B and C (columns from left
to right) at time $3.075$\,Myr. The top row shows the logarithm of the gas
temperature in plane $z=0$; the middle row is the column density integrated
along the $z$-direction; and the bottom row presents the gas radial velocity
relative to the cluster center, green-black corresponds to inwards velocity,
red-white is the outwards velocity, the cyan line separates regions with
subsonic and supersonic velocity. Sink particles are shown in the middle
row by white circles with black borders, and sizes representing their masses as
given in the legend.}
\label{fig:sim:he1}
\end{figure*}

\begin{figure*}
\includegraphics[width=1.0\textwidth]{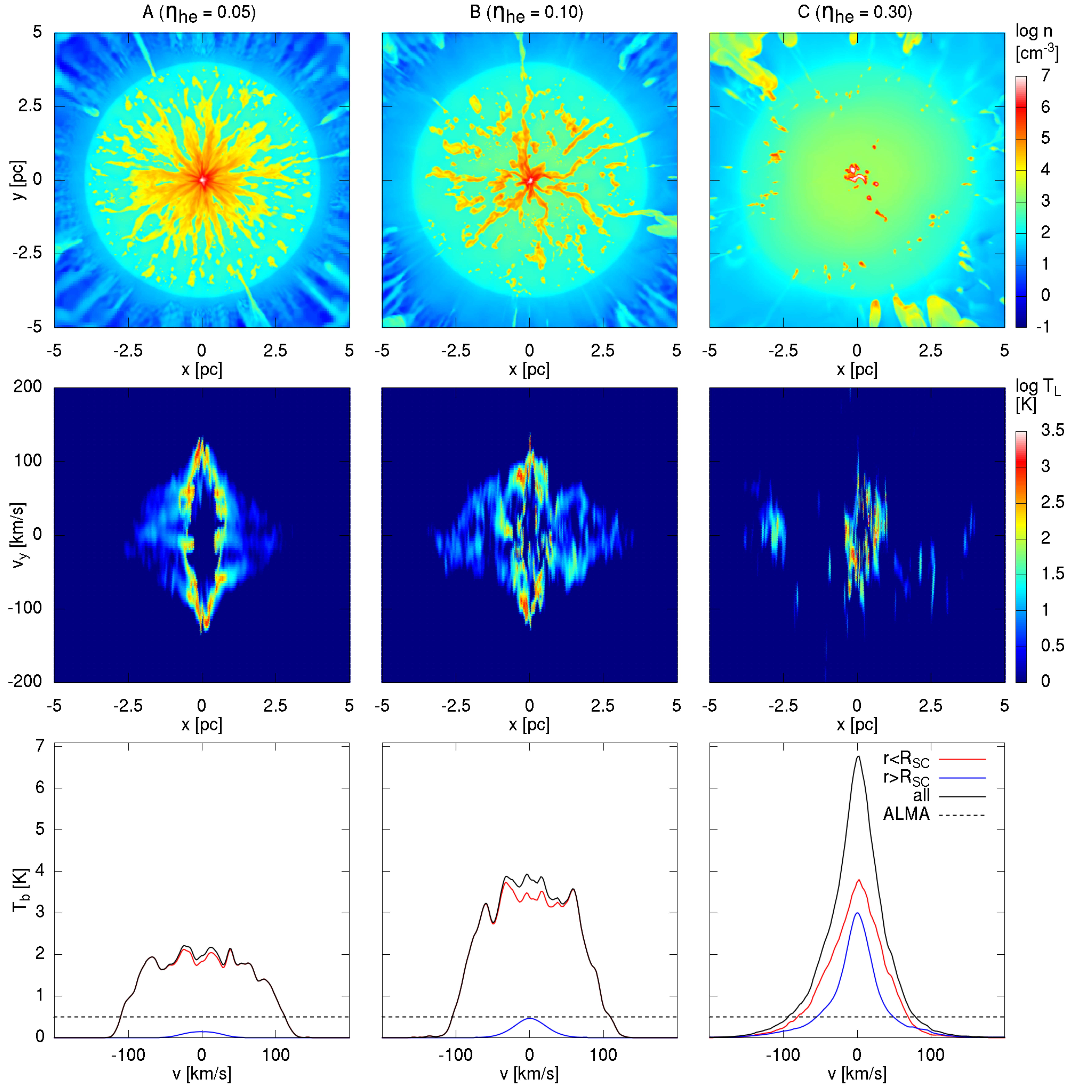}
\caption{Comparison of emission line synthetic observations for models A, B and
C (columns from left to right) at time $3.075$\,Myr. The top row shows the
logarithm of the gas particle density in plane $z=0$; the middle row shows the
position-velocity ($x-v_y$) diagram of the synthetic recombination line emission
from the $z=0$ plane only; and the bottom row presents synthetic line emission
integrated over the whole computational domain. The line profile is calculated
as seen from the $x$-direction and the total emission (black) is split into a
part coming from within the cluster (red) and from outside of it (blue). The
horizontal dashed line ($T_\mathrm{b} = 0.5$\,K) shows a rough estimate of the
ALMA sensitivity with configuration described at the end of \S\ref{ssec:sim}.
}
\label{fig:sim:he2}
\end{figure*}

\begin{figure*}
\plotone{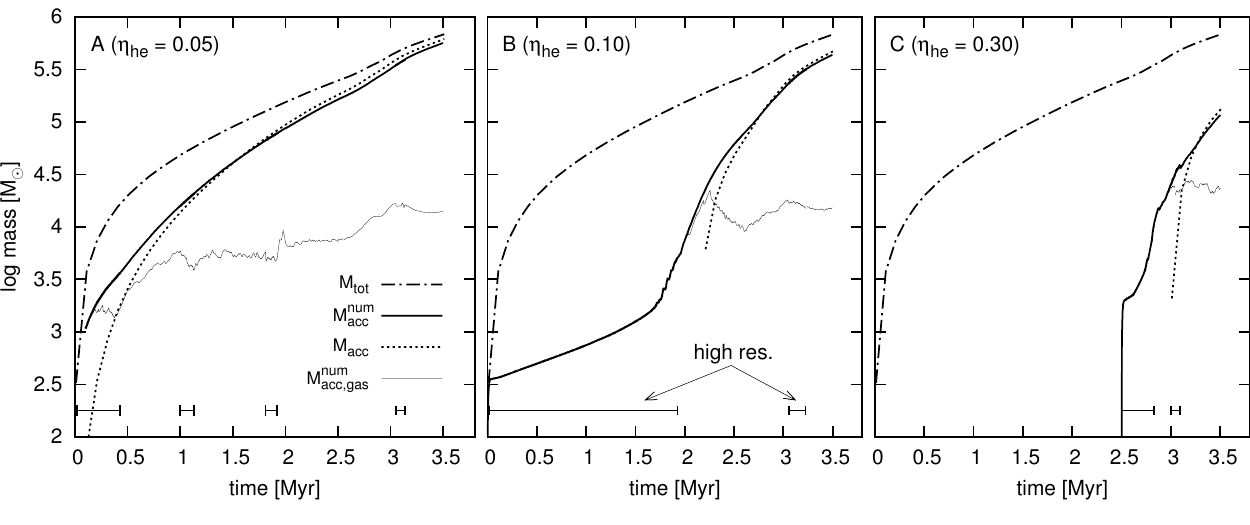}
\caption{Evolution of the inserted and accumulated mass for models A, B and C
(panels left to right). The thick dash-dotted line shows the total amount of
gas, $M_\mathrm{tot}$, inserted to the cluster up to a given time (both winds
and mass loading). The thick solid black line denotes mass accumulated in RHD
simulation, $M_\mathrm{acc}^\mathrm{num}$ (both gas and sink particles), and the
thick dotted is the corresponding accumulated mass, $M_\mathrm{acc}$, estimated
by the semi-analytic model. The thin solid line shows the mass,
$M_\mathrm{acc,gas}^\mathrm{num}$, of the warm and cold gas present in the
simulation computational domain, \ie the accumulated mass without sinks. The
horizontal lines with T-shaped heads at the bottom of the figures denote periods
during which the simulations were calculated at higher resolution (AMR, up to
refinement level 7 corresponding to $512^3$ maximum resolution).
}
\label{fig:mevol}
\end{figure*}

\begin{figure}
\plotone{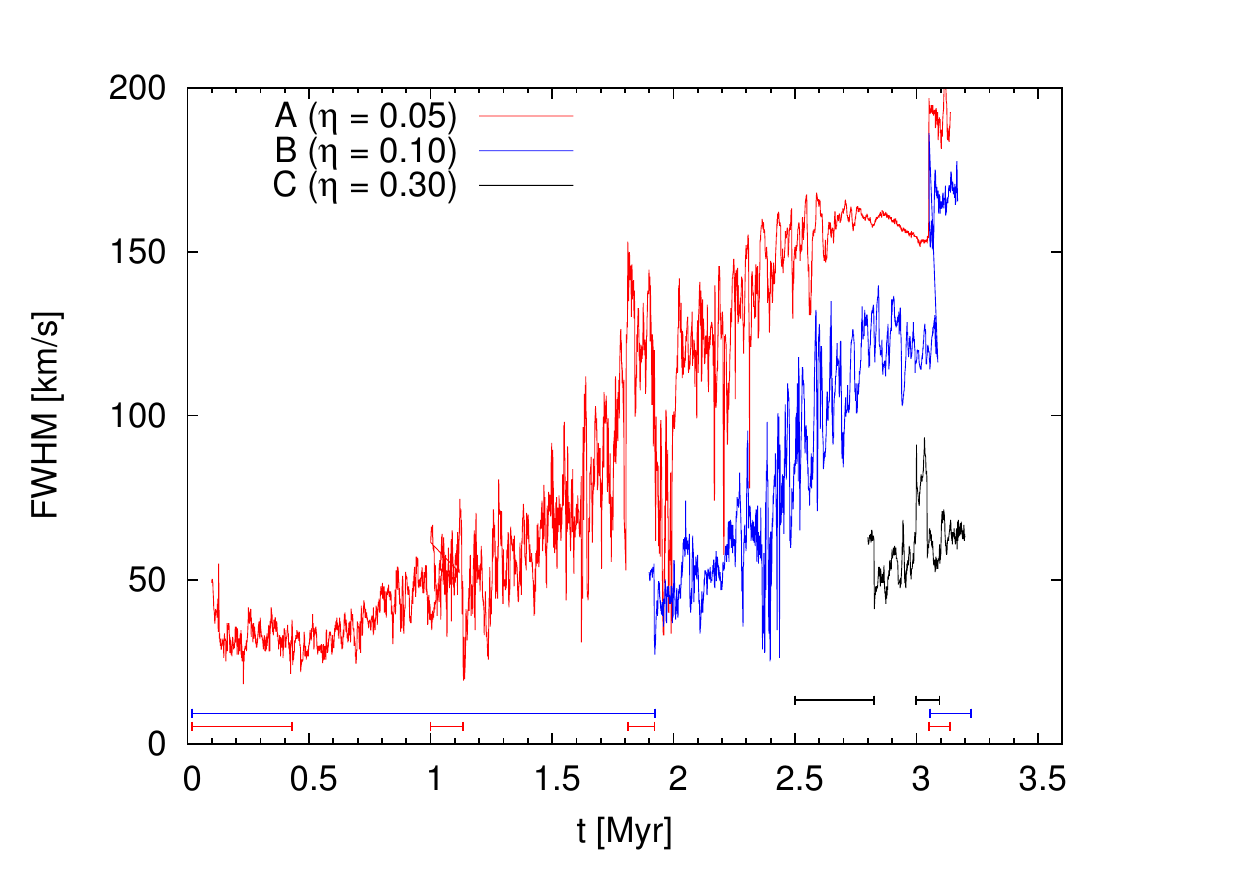}
\caption{Evolution of the synthetic line widths (FWHM) for models A (red), B
(blue) and C (black). The horizontal lines with T-shaped heads and a
corresponding color at the bottom denote periods during which the simulations
were calculated at higher resolution}
\label{fig:FWHM:time}
\end{figure}

\begin{figure}
\plotone{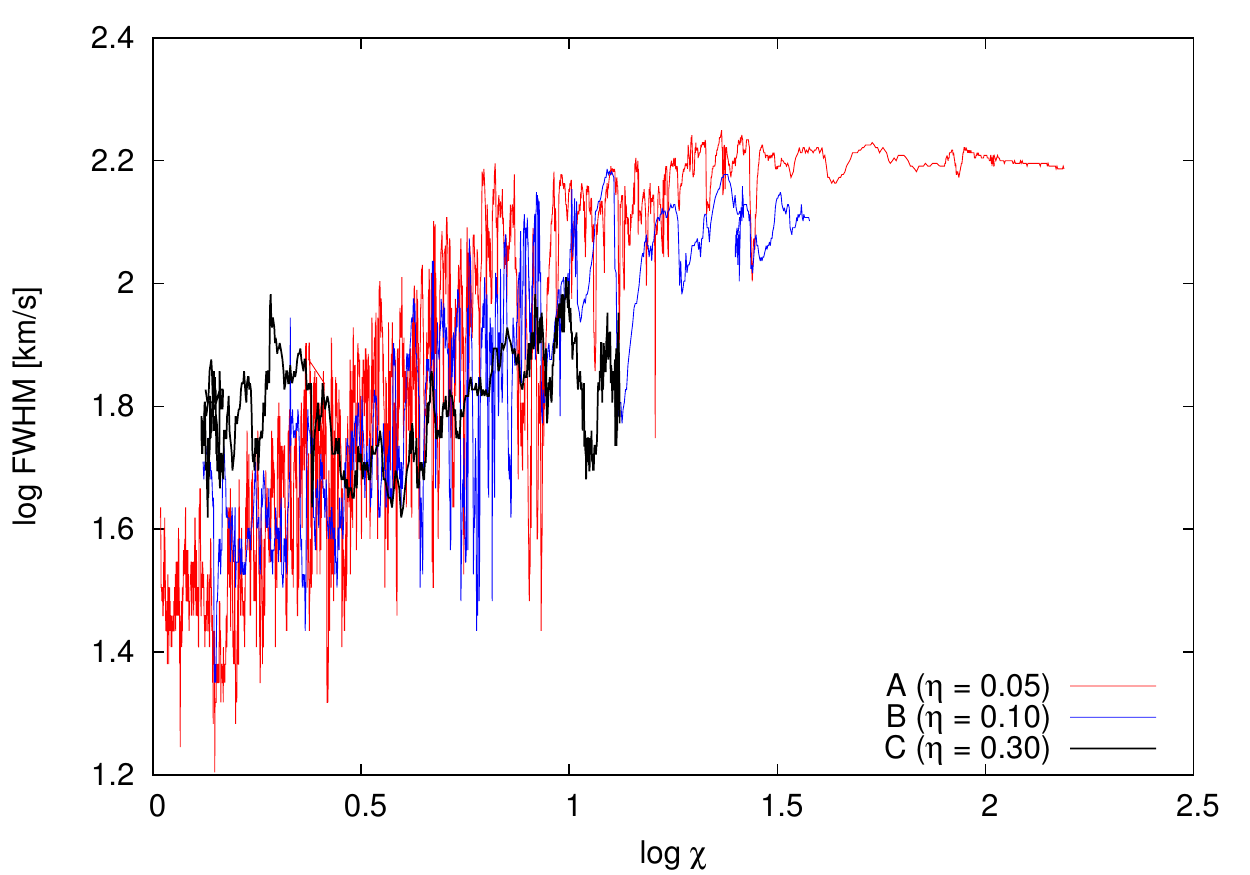}
\caption{The synthetic line widths (FWHM) for models A (red), B
(blue) and C (black) as a function of $\chi$. }
\label{fig:FWHM:chi}
\end{figure}



The behavior of models is governed mainly by the ratio $\chi \equiv
L_\mathrm{SC}/L_\mathrm{crit}$. If it is smaller than $1$, the stationary solution
exists and simulations exhibit a spherically symmetric quasi-stationary
distribution of the wind quantities, in a nearly perfect agreement with the
semi-analytic code. The wind reaches the sound speed at $r = R_\mathrm{SC}$ and
at a certain radius, $R_\mathrm{cool} > R_\mathrm{SC}$, the wind cools down to
temperatures $\sim 10^4$~K. The radius $R_\mathrm{cool}$ decreases with increasing
$\chi$,  approaching $R_\mathrm{SC}$ from the outside.

In models for which $\chi$ exceeds $1$, at time defined as $t_\mathrm{bs}$ (see
Table~\ref{tab:sims} for $t_\mathrm{bs}$ for individual models), clumps start to
form rapidly inside the cluster. All models exhibit one of the two qualitatively
different behaviors illustrated by Figure~\ref{fig:sim:time} for model~A. The
four panels in each row display (from left to right) the particle density in the
plane intersecting the cluster center, the temperature in the same plane, the
column density of gas and the radial velocity in the central plane. The top row
is for time $0.1$\,Myr when $\chi$ just exceeds $1$ and the simulation shows
individual clumps that are either falling into the center, or leaving the
cluster at its periphery. The bottom row, made at $1$\,Myr when $\chi
\approx 2$, shows continuous streams of warm gas that are flowing into the
central region. There, the central clump includes the cold core in which sink
particles are formed.

Figure~\ref{fig:sim:he1} compares the behavior of models A, B and C at
$3.075$\,Myr when $\chi>1$ for all models and they have all accumulated a
substantial amount of gas. On the temperature plots (top row) we see that the
low heating efficiency (model A) leads to a lower temperature of the hot gas and
to a smaller radius at which the wind cools from hot to warm. Further we see
that the cold gas (i.e. shielded regions) exists only in the center in
model~A. In model~B, vast majority of the cold gas is located also in the
cluster center, however, self-shielding occurs scarcely also at higher radii. In
model C, small clumps throughout the whole cluster become ordinarily
self-shielding and cold in their interiors. This is in a good agreement with the
prediction of the semi-analytic model (cf. to Figures~\ref{fig:Mc} and
\ref{fig:Ms}). The middle row of Figure~\ref{fig:sim:he1} (gas column density)
shows that the low heating efficiency (model A) results in a higher volume
filling factor of warm gas occurring in radially inflowing streams, while the
higher heating efficiency model C leads to a lower volume filling factor of warm
gas in chaotically distributed dense clumps. The radial velocity maps (bottom
row of Figure~\ref{fig:sim:he1}) show  the dense gas concentrated in streams (in
models~A and B) flowing inwards from almost all the cluster volume. On the other
hand, clumps in model C have inward velocities at small radii and outward
velocities at larger radii. This is in agreement with the semi-analytic model
that assumes that clumps formed below $R_\mathrm{esc}$ fall to the center while
those forming above $R_\mathrm{esc}$ flow out of the cluster. The escape radii
are, at a given time, $R_\mathrm{esc} = 5.2$, $4.0$ and $2.4$\,pc, for models~A,
B and C, respectively.

The distribution of sink particles, shown together with the column density in
the middle row panels of Figure~\ref{fig:sim:he1}, follows closely the
distribution of the cold gas. In models~A and B with the cold gas only in the
center, a single very massive ($\sim 10^5$\,M$_\odot$) particle is formed
accreting onto itself all cold gas\footnote{Initially, several sinks are formed
in model A, however, all except one are ejected by dynamical interactions with
the most massive sink and infalling gaseous clumps.}. On the other hand, in model
C, several tens of sink particles are formed with masses $10^2 -
10^3$\,M$_\odot$ distributed throughout the cluster volume. As stated before, the
simulations are unable to resolve individual stellar masses and therefore  sink
particles represent clusters of stars or stellar associations rather than
individual objects. Due to the extremely simplified physics of star formation,
sink particles are here regarded as tracers of the star formation location and
their total mass as an upper limit to the mass of second generation stars.

Figure~\ref{fig:mevol} shows the total amount of  mass accumulated inside the
clusters, $M_\mathrm{acc}^\mathrm{num}$ (thick solid line), according to
the RHD simulations, as a function of time for models~A, B and C. For
almost all the time, except a short period after the first sink formation, the
accumulated mass is dominated by the mass of sink particles as seen by
comparing with the accumulated gas only, $M_\mathrm{acc,gas}^\mathrm{num}$ (thin
solid line). Further, $M_\mathrm{acc}^\mathrm{num}$ is compared to the total
amount of mass inserted into the cluster by stellar winds, $M_\mathrm{tot}(t) =
\int_0^t (1+q_m)\dot{M}_\mathrm{SC}(t')dt'$ (dash-dotted line), and to the
semi-analytic estimate of the accumulated mass, $M_\mathrm{acc}$ (dotted line,
Equation~\ref{eq:macc}), derived from semi-analytic calculations.
The above quantities at the end of simulations are also given by
Table~\ref{tab:sims}. We can see that $M_\mathrm{acc}^\mathrm{num}$ is a strong
function of $\eta_\mathrm{he}$: in model A, the majority of the inserted mass
stays in the cluster, while in model C, the fraction of accumulated mass is less
than 20\%. We can also see (both from Table~\ref{tab:sims} and
Figure~\ref{fig:mevol}) that the semi-analytic estimates of the accumulated gas
are very close to the values obtained from the simulations.

\subsection{Observational predictions}
\label{ssec:obspred}

Synthetic spectra calculated for models~A, B and C at time $3.075$\,Myr are
presented in Figure~\ref{fig:sim:he2}. The figure shows the particle density in
the $z=0$ plane (top row), the H30$\alpha$ line emission coming from the same
plane in the position-velocity diagram (middle row; Equation~\ref{eq:TL}) and
the brightness temperature velocity profile $T_b$ integrated over all grid cells
calculated for a virtual telescope with angular resolution $D_\mathrm{res} =
10$\,pc. Note that the line profiles of the three simulated models are
considerably different. Model~A exhibits a broad ($\mathrm{FWHM} \sim
200$\,km\,s$^{-1}$) line with flat and nodulated top. A comparison with the
particle density map and the position-velocity diagram shows that the majority
of the emission comes from the central region with many dense warm inflowing
streams. The highest velocity of the emitting gas occurs close to the center
where the streams are accelerated to velocity $\sim 100$\,km\,s$^{-1}$ by the
gravitational field of the cluster and the central sink particle. In the very
center, the infalling gas becomes shielded and cold and stops to emit in
the recombination line. In this way, the radius at which  self-shielding occurs
determines the FWHM of the line. On the other hand, the line profile of model~C
is much narrower ($\mathrm{FWHM} \sim 100$\,km\,s$^{-1}$) and it has a sharp
peak. Here the emission arises from contributions of many small dense warm
clumps with both inwards and outwards velocities (see the bottom panel of
Figure~\ref{fig:sim:he1}). The line profile of model~B is a transition between
the other two cases, although it seems  qualitatively closer to model~A. A
decomposition of $T_b$ into emission coming from $r<R_\mathrm{SC}$ and from $r >
R_\mathrm{SC}$ (red and blue curves, respectively) shows that the majority of
emission comes from within the cluster in models A and B and that the emission
from both regions is comparable in model~C.

Figure~\ref{fig:FWHM:time} shows the evolution of the FWHM of the synthetic
H30$\alpha$ lines for the three calculated models. One can appreciate two
general trends: (i) the line width grows with time for all models, and (ii) the
line width decreases with increasing $\eta_\mathrm{he}$. All FWHM curves also
exhibit sudden growths at times when the resolution increases and sudden drops
at times when the resolution decreases. This can be understood as a consequence
of the cold regions not being properly resolved: the majority of high velocity
emission comes from the dense warm gas flowing into the cold regions. The higher
resolution leads to smaller cold regions with the inflowing warm gas reaching
higher densities and velocities. This implies  that the line profiles and
widths, regardless of the amount of accumulated mass, are not well resolved
with the numerical resolution used and therefore should be taken just  as
indicators of general trends and not to draw quantitative predictions.

Figure~\ref{fig:FWHM:chi} shows the FWHM of the synthetic lines as a function of
$\chi$. Models~A and B show a very similar behavior: the FWHM grows with $\chi$
monotonically for $\log(\chi)$ between $0$ and $\sim 1.2$ and stays
approximately constant for higher values. Since in both models the dense warm
gas occurs mainly in the inflowing streams, we interpret it so that the FWHM is
given by the maximum velocity of streams. Greater value of $\chi$ leads to
higher $R_\mathrm{esc}$, and therefore streams inflow into the center from
larger radii and with higher velocity leading to the growth of FWHM with $\chi$.
For $\log(\chi) \gtrsim 1.2$ the outer boundaries of the streams reach almost
the cluster border, and therefore their growth is not further possible and
FWHM$(\chi)$ saturates. On the other hand, model~C does not show a clear
FWHM$(\chi)$ dependence. This most likely is because the emission in this model
comes mainly from individual clumps formed at various radii having more random
(both inwards and outwards) velocities.

Finally, we estimate the observability of the warm gas predicted by the
simulations. The closest known objects with comparable parameters (mass, radius,
age) are super star clusters in the interacting galaxies NGC4038/9 (Antennae).
At their approximate distance $\sim 20$\,Mpc \citep{2008AJ....136.1482S} and
diameters $\sim 10$\,pc, the correponding angular resolution is
$D_\mathrm{res} \sim 0.1$". Using the ALMA sensitivity calculator, we estimate
the integration time needed to reach sensitivity $0.5$\,K (shown as dashed
horizontal lines in the bottom panels of Figure~\ref{fig:sim:he2}) with $40$ 12m
antennae and bandwidth $200$\,km\,s$^{-1}$ to be $18$\,mins. Therefore, we
conclude that it should be in principle possible to test the presented model
with  observations using the appropriate ALMA configurations.


\subsection{Parameter space study}
\label{ssec:parspace}


\begin{figure}
\plotone{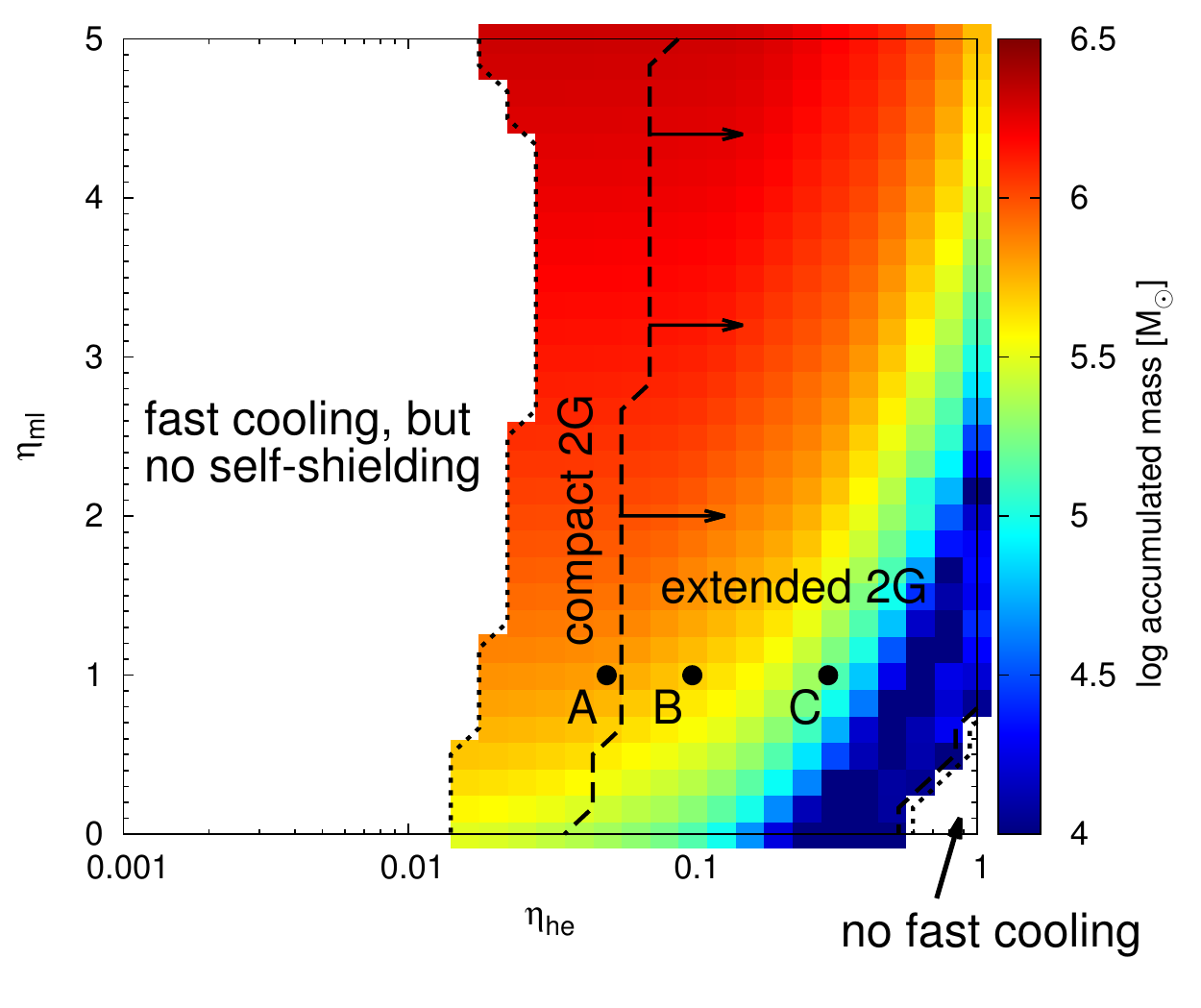}
\caption{Map of the $\eta_\mathrm{he}$-$\eta_\mathrm{ml}$ parameter space for
models with parameters given by Table~\ref{tab:compar}. The color shows the
amount of mass accumulated during first 3.5 Myr for a model with given
parameters. The dotted and dashed lines separate four qualitatively different
regions described in the text (see \S\ref{ssec:parspace}). Location
of models A, B and C is shown by black circles. Horizontal arrows attached
to the dashed vertical line denote that it should be interpreted as a lower limit.
}
\label{fig:heml}
\end{figure}

Motivated by the excellent agreement between the accumulated mass in numerical
models $M_\mathrm{acc}^\mathrm{num}$ and the semi-analytic estimate
$M_\mathrm{acc}$ we consider a larger subset of the parameter space
$\eta_\mathrm{he}$-$\eta_\mathrm{ml}$ using the semi-analytic code. We calculate
a grid of models with parameters given in Table~\ref{tab:compar} (which
correspond to our numerical models A, B and C) and vary the heating efficiency
and the mass loading in intervals $\eta_\mathrm{he} \in (0.001,1)$ and
$\eta_\mathrm{ml} \in (0,5)$. For each model we calculate the first $3.5$\,Myr
of the evolution and evaluate: (i) whether rapid cooling occurs during that
period (i.e. $\chi > 1$), (ii) whether the central clump becomes
self-shielding (i.e. $M_\mathrm{c} \equiv M_\mathrm{acc} >
\min(M_\mathrm{max,sh}, M_\mathrm{c,sh})$, see Equations~\ref{eq:macc},
\ref{eq:mcsh} and \ref{eq:Mmaxsh}), (iii) whether the infalling stream becomes
self-shielding (i.e. $M_\mathrm{s} > \min(M_\mathrm{max,sh},
M_\mathrm{s,sh})$, see Equations~\ref{eq:macc}, \ref{eq:ms} and \ref{eq:mssh}),
and (iv) if self-shielding occurs, what is the amount of accumulated gas
$M_\mathrm{acc}$ (Equation~\ref{eq:macc}).

The resulting map of the $\eta_\mathrm{he}$-$\eta_\mathrm{ml}$ parameter space
is shown in Figure~\ref{fig:heml}. The color represents the amount of
accumulated gas, $M_\mathrm{acc}$, and it is plotted only if self-shielding
occurs (otherwise, it is left white). We can identify four qualitatively
different regions. First, a small region with high $\eta_\mathrm{he}$ and small
$\eta_\mathrm{ml}$ (bottom right) where $\chi$ never exceed $1$ and
rapid cooling does not occur. Second, the region with $\eta_\mathrm{he}
\lesssim 0.02$ where even though rapid cooling occurs inside the
cluster, the gas never achieves self-shielding. This is because low
$\eta_\mathrm{he}$ leads to relatively low pressure and hence to a low warm gas
density, and therefore all the warm gas that can be accommodated inside the
cluster is ionized by the EUV radiation from the stars. Consequently, in these
two regions secondary star formation is inhibited. The third region lies
between $\eta_\mathrm{he} \sim 0.02$ and $0.05-0.08$ (marked "compact 2G",
includes model~A); here rapid cooling occurs and the central clump is
able to self-shield. However, due to the relatively low density of the warm gas,
the infalling streams are never dense and massive enough to reach
self-shielding. In this case, secondary star formation becomes possible only in
the central clump. In the fourth region with $0.05-0.08 < \eta_\mathrm{he} <
0.8-1$ (marked extended 2G, includes models~B and C), both the central clump and
the infalling streams achieve self-shielding conditions and thus secondary star
formation becomes  possible in both the central clump and the infalling streams.
The vertical dashed line separating the third and the fourth region represents
only a lower limit on $\eta_\mathrm{HE}$ and the more realistic border between
those two regions lies probably slightly rightwards (see below). This is
indicated by horizontal arrows attached to the vertical dashed line.

Numerical models A, B and C are denoted by black circles in
Figure~\ref{fig:heml}. Model A lies in the region predicting a compact and
central 2G sub-cluster formation, in good agreement with the numerical results.
Throughout the whole evolution, only the central clump is
self-shielding, and sink particles form only in the very center. Similarly,
model C lies in the region where the extended 2G sub-cluster should be formed,
and again it is in a good agreement with the model behavior. The gas becomes
self-shielding even at larger radii while  falling into the cluster center, and
a larger number of less massive sink particles form throughout the cluster. On
the other hand, model B lies also in the region of "extended 2G", but its
behavior is closer to that of model A. This implies   that the line separating
"compact 2G" and "extended 2G" regions marks only a lower limit in
$\eta_\mathrm{he}$. The model is classified as the one with an "extended 2G" if
during the calculation  there is a period when $M_\mathrm{s} > M_\mathrm{s,sh}$.
As this period can be arbitrarily short, the majority of stars can still form in
the central clump. Moreover, the criterion only evaluates whether self-shielding
in streams occur, but the semi-analytic model is unable to calculate whether the
cold gas collapses into stars / sink particles. The top middle panel of
Figure~\ref{fig:sim:he1} indeed shows that self-shielding can rarely and
marginally occur also at large radii. Therefore we conclude that the more
realistic estimate of the location of the line separating "compact 2G" and
"extended 2G" would be somewhere between the calculated line and the position of
model~C, i.e. approximately at $\eta_\mathrm{he} = 0.08 - 0.3$.


\section{Discussion}
\label{sec:discussion}


The model used in this work includes many simplifications and caveats.
Here we list the ones that we consider most important. One of the most serious
problems is probably the unknown origin of the low heating efficiency treated as
a free parameter. In principle, it can be any type of additional cooling, not
accounted for by the gas cooling included in our model. One possibility could be
cooling at the transition layer between the hot gas and warm clumps combined
with the thermal conduction transporting the heat from the hot gas onto clump
surfaces. We plan to explore this option in future work. Inclusion of the
thermal conduction leading to evaporation of pre-existing clumps may also
provide self-consistent mechanism for mass loading, which is here also treated
as a free parameter. Furthermore, the physics of gas with temperatures below
$10^4$\,K and 2G star formation process is extremely simplified in the numerical
code and not present in the semi-analytic model. As a result, the 2G mass is
probably overestimated and the provided values should be contemplated rather as
upper limits. Feedback from the 2G stars is also missing, even though it
probably behaves in a similar way as feedback from 1G stars, and can be
considered as a local enhancement of mass and energy deposition rates within the
framework of the model. Another simplification is the distribution of wind mass
and energy, in our model evenly supplied within the cluster volume, assuming
that the wind-wind collisions redistribute smoothly the mass and mechanical
energy of the winds. This has been well justified by \citet{2000ApJ...536..896C,
2006MNRAS.369..860C} for adiabatic models, however, the applicability of this
approach in case of more complex physics including cooling and radiation is less
clear.

The presented model predicts the formation of 2G stars out of stellar winds from
the 1G. Such models have already been suggested e.g. by
\citet{2007A&A...464.1029D} and \citet{2010MNRAS.407..854D}. However, our model exhibits two
unique features. Firstly, it predicts that even fast stellar winds with
velocities exceeding thousands of km/s can be captured inside the cluster and 2G
stars can form out of them, while all  previous models assumed that stellar
winds have to be slow \citep[e.g.][]{2008MNRAS.391..825D} in order to contribute
to 2G star formation. Moreover, the model makes a clear link between the cluster
global properties and the secondary star formation by predicting that it can
occur only if the cluster is massive and compact enough (i.e. $L_\mathrm{SC} >
L_\mathrm{crit}$). Another unique feature is that our model provides a
self-consistent mechanism predicting that 2G stars form in a small central part
of the cluster if the heating efficiency is small. The existing models
\citep[see e.g.][]{2013A&A...552A.121K} sometimes assume that 2G stars can form
in the cluster central region, because they form out of massive stars located
close to the center due to primordial mass segregation. However, it is only an
assumption and the hypothesis of primordial mass segregation has been questioned
by recent observations of NGC3603 with VLT/SPHERE using extreme adaptive optics
\citep{2016A&A...588L...7K}.

The model predicts secondary star formation occurring in young massive
clusters with solar metallicity, and it therefore naturally raises the  question of
whether it could be tested by observations of nearby young massive clusters.
Photometric observations of intermediate age \citep{2003AJ....125..770B,
2008ApJ...681L..17M, 2009A&A...497..755M} and recently even young
\citep{2015MNRAS.450.3750M, 2016MNRAS.458.4368M} massive clusters in the LMC indeed
suggest the presence of multiple stellar populations. However, other explanations
of multiple episodes of star formation have been also suggested and recently
\citet{2016MNRAS.458.4368M, 2016MNRAS.460L..20B} argue that the effect of 
stellar rotation may provide the most plausible one. More promising can be the 
detection of warm dense gas with a high velocity dispersion as suggested in this
work. The emission lines showing the presence of the warm gas have been observed
in embedded clusters in galaxies as NGC~5253 \citep{2012ApJ...755...59B,
2015Natur.519..331T, 2016ApJ...823...38S}, NGC~4449 \citep{2015AJ....149..115S}
and in Antennae interacting galaxies \citep{2007ApJ...668..168G}. The predicted
emission line profiles seems to show a significant differences between the emission of HII
regions and the emission from the cluster interior. A more detailed comparison of
the predicted versus observed profiles should be performed in the future.

An interesting question is whether the rapidly cooling winds model could explain
the origin of multiple stellar populations observed commonly in globular
clusters. Formation of globular clusters is a complex field and even though many
mechanisms have been suggested, a fully satisfactory model does not seem to
exist \citep{2015arXiv151001330B}. We have here described a basic mechanism that
always leads to rapid cooling and mass accumulation providing there is enough hot
and relatively dense gas inside the cluster. Therefore, a critical question
determining whether the model could work depends on whether stellar evolution
models for low metallicity massive stars predict winds with a large enough mass
loss. One possibility could be models of fast rotating massive stars
\citep{2007A&A...464.1029D} or massive binaries \citep{2009A&A...507L...1D,
2016ApJ...825..118T}. The presented model differs from the other mentioned
models by the fact that the wind always goes through the hot phase and
contributions from various types of stars and the pristine (mass loaded) gas mix
completely together. This, on the one hand, could  explain the presence of Li
(signature of pristine gas) in 2G stars, on the other hand, it does not seem
straight forward  to explain the  extreme abundance patterns (e.g.
the high oxygen depletion), as it only happens in some (very massive) types of
stars. A feature in favor of the rapidly cooling winds model is the mentioned
self-consistent mechanism predicting the formation of 2G stars in the very
center, which  provides an ideal setup for the removal of 1G stars by
combination of gas expulsion and tidal forces as described in
\citet{2015MNRAS.452..924K}. Another attractive feature of our model is the link
between secondary star formation and the global parameters of the cluster (mass,
radius, metallicity, \dots), which provides a natural explanation as to why only
globular clusters (or in general massive cluster), and not less massive open
cluster or field stars, exhibit features related to multiple stellar
populations.


\section{Conclusions}
\label{sec:conclusions}


We have studied a model of \emph{rapidly cooling shocked stellar winds}
in young massive clusters and estimate the circumstances under which secondary
star formation, out of the reinserted  winds from a first stellar generation is
possible. We have used two implementations of the model: a highly idealized
computationally inexpensive spherically symmetric semi-analytic model, and a
complex three-dimensional radiation-hydrodynamic simulations. The model
determines whether the hot shocked stellar winds inside the cluster become
thermally unstable and form dense clumps, whether these clumps self-shield
against the stellar EUV radiation and cool further where ever it may happen. The
model also determines the fraction of stellar wind mass that cools down and
feeds secondary star formation. Both implementations show a good agreement for
the three calculations made with different values of the heating efficiency
of the shocked stellar winds. Further, we have used the semi-analytic model to
explore a subset of the parameter space covering a wide range  of the
observationally poorly constrained parameters: the heating efficiency,
$\eta_\mathrm{he}$, and the mass loading, $\eta_\mathrm{ml}$. Finally, we have
calculated the emission in the H30$\alpha$ recombination line, analyzed its
velocity profile and estimated its intensity for super star clusters at the
distance of the interacting galaxies NGC4038/9 (Antennae).

Our conclusions are as follows:

\begin{enumerate}

\item With more accurate and complex numerical model including gravity and
ionizing radiation we confirm our previous findings \citep{2005ApJ...628L..13T,
2008ApJ...683..683W} that in  young, massive and compact clusters, the resultant
thermalized shocked stellar winds become  thermally unstable. This leads to the
formation of dense warm clumps before leaving the cluster volume while composing
a cluster wind. The dense clumps cool further as they self-shield themselves
from the EUV radiation, triggering then the formation of next generations of
stars. In this way, the reinserted stellar wind material, expected in adiabatic
calculations to be expelled from the cluster volume with velocities largely
exceeding the escape velocity of the cluster, can be captured and used for
secondary star formation.

\item The fraction of the mass reinserted through 1G stellar winds which
accumulates inside the cluster and becomes available for secondary star
formation is a function of cluster parameters, and it can be large ($>50$\%) for
sets of reasonable parameter. Specifically, for clusters with 1G stellar mass
$\sim 10^7$\,M$_\odot$, half-mass radius $2.38$\,pc, mass loading
$\eta_\mathrm{ml} = 1$ and heating efficiencies $0.05$, $0.1$ and $0.3$, the
fractions are $88$\,\%, $68$\,\% and $17$\,\%, respectively. The corresponding
masses of gas available for  secondary star formation are $6.1\times 10^{5}$,
$4.7\times 10^{5}$ and $1.2\times 10^{5}$\,M$_\odot$. Thus our model 
suffers also  the "mass budget problem" encountered  in former scenarios
trying to explain multiple populations observed in globular clusters: the  mass
fraction of the second stellar generation is too low unless a
substantial fraction of 1G stars are later removed from the cluster.

\item The presented model provides a self-consistent mechanism predicting
the formation of 2G stars only in the central zones of the cluster. The crucial
parameter determining where the 2G stars form is the heating efficiency: if it
is low (of order $1-10$\%), 2G stars form only in the center; if it is larger,
2G stars form everywhere throughout the cluster (see regions "compact 2G" and
"extended 2G" on the parameter space map in Figure~\ref{fig:heml}). The heating
efficiency is closely related to the temperature of the hot shocked wind within
the cluster and there is some observational evidence, that it may indeed be low
\citep{2009ApJ...700..931S, 2014MNRAS.442.2701R}. This is interesting in terms
of the aforementioned mass budget problem, because if 1G and 2G stars are
spatially separated in this way, a substantial fraction of 1G stars can be lost
due to the primordial gas expulsion and the subsequent dynamical evolution
\citep{2016MNRAS.457..479K}.

\item The model predicts that a cluster with studied parameters and age
$2-3$\,Myr should contain in its interior a dense warm gas in amounts of the
order of $10^4$\,M$_\odot$. This gas can be traced e.g. by observing hydrogen
recombination lines. The line widths predicted by the model are in an
approximate agreement with observations of Br$\gamma$ line for super star
clusters in Antennae galaxies \citep{2007ApJ...668..168G}. The intensities of
the H30$\alpha$ radio recombination line calculated for the modelled cluster at
the distance of Antennae should make the warm gas detectable with the convenient
configuration of ALMA at reasonable integration times.

\end{enumerate}


\begin{acknowledgments} 

We thank the anonymous referee for constructive and valuable comments. We thank
Dorottya Sz\'ecsi for careful reading of the text and her useful suggestions.
This study has been supported by project 15-06012S of the Czech Science
Foundation, the institutional project RVO:67985815, the CONACYT-M\'exico
research grant 167169, the Bilateral agreement between CONACYT and the Czech
Academy of Sciences grant 17048, and the ISSI project "Massive star clusters
across the Hubble time". This work was supported by The Ministry of Education,
Youth and Sports from the Large Infrastructures for Research, Experimental
Development and Innovations project "IT4Innovations National Supercomputing
Center – LM2015070".

\end{acknowledgments}

\bibliographystyle{aasjournal}
\bibliography{coolwinds2G}

\end{document}